\documentclass[twocolumn]{aastex631}

\graphicspath{{./}{figures/}}
\usepackage{amsmath}
\usepackage{bm}
\usepackage{color}
\usepackage{hyperref}
\makeatletter
\DeclareRobustCommand{\rvdots}{
  \vbox{
    \baselineskip4\p@\lineskiplimit\z@
    \kern-\p@
    \hbox{.}\hbox{.}\hbox{.}
  }}
\makeatother
\begin{document}
\newcommand{\dm}{\mathrm{DM}}
\newcommand{\snr}{\mathrm{S/N}}
\newcommand{\sfr}{\mathrm{SFH}}

\title{A Comprehensive Study of the Energy and Redshift Distributions of the Fast Radio Burst Population Based on the First CHIME/FRB Catalog}

\correspondingauthor{Can-Min Deng}
\email{dengcm@gxu.edu.cn}
\author{Qing-Zhen Lei}
\affiliation{ Guangxi Key Laboratory for Relativistic Astrophysics, Department of Physics, Guangxi University, Nanning 530004,China}
\author{Xin-Zhe Wang}
\affiliation{ Beijing Normal University, School of Physics and Astronomy, Beijing 100875, China.}          
\author{Can-Min Deng}
\affiliation{ Guangxi Key Laboratory for Relativistic Astrophysics, Department of Physics, Guangxi University, Nanning 530004,China}

\begin{abstract}
Fast radio bursts (FRBs) are brief, high-energy bursts of radio waves from extragalactic sources, and their origin remains an open question. In this paper, we perform a comprehensive analysis of the FRB population using the first CHIME/FRB catalog, focusing on their energy  and redshift  distribution, with careful consideration of selection effects. We investigate a range of models, including the Schechter function and the broken power-law function for the energy distribution, and several redshift evolution models, such as the star formation history (SFH) model, as well as models incorporating time delays relative to the SFH or additional redshift evolution factors. Our results indicate that the energy distribution of FRBs is best described by the Schechter function, with a power-law index of $\gamma = -1.49^{+0.37}_{-0.27}$ and a characteristic cutoff energy of $E_\mathrm{c} = 2.82^{+2.43}_{-1.47} \times 10^{41}$ erg. Furthermore, we find no evidence for redshift evolution in the energy distribution of FRBs. In terms of their redshift distribution, our analysis shows that it follows the cosmic SFH, without requiring additional delayed components or redshift evolution factors, suggesting that most FRBs likely originate from young stellar populations. Simultaneously, we infer a local volumetric rate of $\Phi_0 = 4.68^{+4.66}_{-2.39} \times 10^{4} \rm \ Gpc^{-3}yr^{-1}$ for $E>10^{39}$ erg. These results, robust against CHIME observational biases, may provide new insights into the underlying properties of the FRB population.

\end{abstract}


\section{Introduction \label{sec:intro}} 
Fast radio bursts (FRBs) are brief but powerful extragalactic radio pulses, typically lasting only a few milliseconds \citep{lorimer+2007, thornton+2013, cordeschatterjee+2019, ZhangBing+2023}. Since their discovery over a decade ago, nearly a thousand FRB sources have been detected \citep{xu+2023}. Among these, a small fraction repeat, while the majority exhibit only a single burst. However, the question of whether these two types of bursts are fundamentally different remains an ongoing debate \citep{2019MNRAS.484.5500C, 2021A&A...647A..30G, cui+2021, catalog1, Zhong+2022, Chen+2022, 2023PASA...40...57J}. A variety of physical models have been proposed to explain the origins of FRBs, with magnetar-based models being the most widely accepted \citep{Metzger+2017, Beloborodov+2017, Margalit+2018, Wadiasingh+2019, Wadiasingh+2020, Beloborodov+2020, mereghetti+2020_sgr1935, Lu+2020, Yang+2021, Chime+2022, Beniamini+2023}. This aligns with observations of some repeating FRB host galaxies, which suggest a connection between FRBs and dwarf galaxies with active star formation \citep{2017ApJ...834L...7T, marcote+2020, 2022Natur.606..873N}. The discovery of  FRB-like bursts further supports the idea that at least some FRBs originate from young magnetars formed by the collapse of massive stars \citep{bochenek+2020_stare2_sgr1935, chimefrb_sgr1935}.
However, the discovery of the repeating FRB 20200120E within a globular cluster in M81 challenges this view, suggesting that some FRBs may arise from older stellar populations \citep{chatterjee+2017_m81, marcote+2017_m81, bassa+2017_m81, marcote+2020, tendulkar+2021_m81,  fong+2021_m81,nimmo+2022_m81, ravi+2022_m81}. This had led to the speculations that this particular FRB could be located in a binary system, which is the most common system in globular clusters \citep{2023Natur.618..484L}.  Theoretically, the accretion-driven binary models have been proposed as alternative sources for active FRBs, offering an alternative to the magnetar-driven model \citep{2021ApJ...922...98D, 2021ApJ...917...13S}. However, no universally accepted model exists, and the physical origin of FRBs remains a mystery.

The study of FRB populations, particularly their energy and redshift distributions, is key to understanding their intrinsic properties and constraining their origins. Analyzing the energy distribution can shed light on the underlying mechanisms, allowing comparisons with other transient sources to explore possible origins. Similarly, the redshift distribution varies across different models, and acquiring the intrinsic redshift distribution helps constrain these models. However, significant challenges persist. Many FRBs lack precise distance measurements, and the dispersion measure (DM) is often used as a proxy \citep{ioka+2003_dmz, inoue+2004_dmz, deng+2014dmz, macquart+2020}, leading to considerable uncertainties, especially for small samples. Moreover, inaccurate localization in most cases results in biased fluence $F$ measurements, further complicating the analysis.
Despite these challenges, a number of studies have attempted to explore the energy and redshift distributions of FRBs using  samples from  radio telescopes like Parkes and ASKAP \citep{luo+2018, 2019JHEAp..23....1D, locatelli+2019_energy, 2019A&A...632A.125G, Lu+2020, hashimoto+2020_energy, Arcus+2021, 2021A&A...651A..63G, 2021MNRAS.501..157Z, james+2022_energy}. However, due to the limited sample size and varying instrumental selection effects, these early results have been unable to effectively constrain the key properties of FRBs, particularly their redshift distributions.

Fortunately, thanks to its wide field of view and high sensitivity, the CHIME telescope has provided a large, uniform sample of FRBs, making it an invaluable resource for studying their energy and redshift distributions \citep{catalog1}. Several studies have already used this sample to investigate the FRB population. For example, \cite{zhang+2022,qiang+2022,zhang+2023} used Monte Carlo simulations and found that the FRB population does not follow the star formation history (SFH) but rather exhibits a significant delay relative to it. Similarly, \cite{2024ApJ...962...73L, 2024ApJ...969..123L} reached similar conclusions using Bayesian inference methods. Meanwhile, \cite{hashimoto+2022, chen+2024_redshift,Zhang+2025} derived a simple power-law form for the redshift distribution of FRBs and observed a strong evolutionary relationship between FRB energy and redshift. However, these studies seem to often overlook CHIME’s complex selection effects and do not  address the bias in fluence measurements.  Consequently, studies that neglect these multidimensional instrumental selection effects may yield biased or unreliable conclusions, and their results should be interpreted with caution. Some studies \citep[e.g.,][]{Gardenier+2021, James+2023, Wang+2024, 2025A&A...698A.127M} have addressed important aspects such as CHIME’s beamshape and one-dimensional DM selection effects.

To  address  for these issues, \citet{shin+2023} developed a robust method designed to handle CHIME's observation biases by using the injection system data \citep{merryfield+2022}, providing a more accurate study of the energy and redshift distributions. Their results indicated no significant evidence that the FRB population deviates from the SFH.
But unfortunately, they did not examine the possibility of other more models, for instance, {they did not examine} whether the frequently discussed delay model is feasible. In addition, as mentioned above, is the redshift distribution possibly of power-law form? Does the energy have redshift evolution? Or is the energy distribution in the form of broken power-law? These questions are worth exploring systematically.

Therefore, in this work, we aim to conduct a comprehensive study  by systematically exploring a variety of energy and redshift models, including models that account for time delays relative to the SFH and those that include potential redshift evolution. We believe that these are crucial for understanding the physical origin of FRBs.

The structure of this paper is arranged as follows. In Section~\ref{sec:data}, we summarize the usage of the dataset. Section~\ref{sec:Methodology} describes the methods we employed. Section~\ref{sec:fdm_model} introduces the various energy and redshift distribution models
of FRBs considered in this paper. The results 
are presented in Section~\ref{sec:RESULT}, followed by a detailed discussion in Section~\ref{sec:Discussion}. Finally, the conclusions are drawn in Section~\ref{sec:Conclusions}

\section{The Data set  \label{sec:data}}
In this work, we used the  CHIME/FRB catalog~1 (hereafter "Catalog~1") data for the study. 
To be precise, the specific data used in our analysis consists of 225 fast radio bursts\footnote{The data products are provided at https://github.com/kaitshin/CHIMEFRB-Cat1-Energy-Dist-Distrs}, which were selected by \citet{shin+2023} from Catalog 1 based on certain criteria to eliminate as much as possible the selection effects that are difficult to remove. This 225 bursts include 210 non-repeating FRBs and 15 repeating sources.

In addition, we also used the data of the injection system \citep{merryfield+2022} to handle the selection effect of CHIME FRBs.  
In the data of this injection system, a total of $\mathrm{N}_\mathrm{inj}=5\times10^6$ FRB events were injected, the fraction of surviving sky locations is $\mathrm{f}_\mathrm{sky}=0.0277$. 96,942 events planned to be injected, 84,697 were successfully injected, with an injection efficiency of $\epsilon_\mathrm{inj}=0.874$. Ultimately, 39,638 events were eventually successfully detected and assigned a bonsai $\snr$.

\section{Methodology \label{sec:Methodology}}
The large, uniform sample provided by CHIME enables a systematic study of the population properties of FRBs. However, to use CHIME/FRB data, it is crucial to correct for the telescope's selection effects effectively. To address this, the CHIME collaboration developed an injection system specifically designed to account for these effects \citep{merryfield+2022}. This system allows for straightforward corrections of selection effects in population studies. It is also important to note that the fluence data in the CHIME/FRB catalog are not calibrated for the beam response due to the imprecise localization of sources. As a result, the fluence values are systematically underestimated and should be treated as lower limits  \citep{catalog1}. Fortunately, for cases with moderate fluence values, the signal-to-noise ratio ($\snr$) serves as a reliable proxy for fluence and can be well measured \citep{merryfield+2022}. In this work, we closely follow the methods described in \citet{shin+2023}, and an overview of this method is provided below.

To infer or constrain the luminosity function and redshift evolution of cosmological sources, it is common practice to simultaneously fit their luminosity and distance information. In our case, this requires constructing the theoretical distribution $R(F,\rm{DM} |\bm\lambda)$ of fluences and DM for FRBs, and then fitting it to the observed data. The process begins with the construction of the likelihood function. Since the goal is to fit the number of FRBs observed in the two-dimensional ($F$, DM) distribution, the logarithmic binned Poisson likelihood is used \citep{Baker+1984, Ivezi+2014, pdg2020}:
\begin{eqnarray} 
\label{eqn:Poisson_likelihood} 
\mathrm{log}\mathcal{L} = \sum_{ij} \left( {n}_{ij} \log(\zeta_{ij}) - \zeta_{ij} - \log({n}_{ij}!) \right),
\end{eqnarray}
{where ${n}_{ij}$ denotes the observed $F$-$\dm$ distribution, with $i$ indexing the $F$ bin and $j$ indexing the $\dm$ bin. $\zeta_{ij}$ represents the model-predicted value for ${n}_{ij}$, which is related to the model parameters $\bm\lambda$. As noted earlier, since beam-corrected $F$ values are not available in the Catalog~1 data, we use S/N as a proxy for $F$. }
The likelihood is evaluated using 15 logarithmically spaced $\snr$ bins ranging from 12 to 200, and 20 logarithmically spaced $\dm$ bins from 100 to 3500 pc cm$^{-3}$, the fitting was performed using the differential  distribution, following \cite{shin+2023}.

The next step is to accurately compute $\zeta_{ij}$. Given the selection effects of the CHIME telescope, it is essential to apply corrections to ensure a reliable population analysis. The most robust approach to achieve this is by leveraging the injection system developed for CHIME FRBs \citep{catalog1, merryfield+2022}. Ideally, one would sample each set of model parameters during the fitting process, inject them into the real-time pipeline, and compare the resulting simulated observations with real data to assess the goodness of fit. Crucially, this method accounts for all CHIME detection biases by measuring the fraction of injected events that are successfully detected. However, since injecting a single set of parameters is highly time-consuming, this approach is impractical. Instead, a more efficient method is adopted.  
By constructing appropriate weights, different candidate population models $\zeta_{ij}$ can be evaluated. These weights, or referred to as scaling factors, are designed to transform the injected population model, which is fixed and determined by CHIME's injection campaign, into any other desired population model. These same weights must then be consistently applied to the detected injections when matching any given model prediction. 
{Each injected burst $m$ within a given bin ${n}_{ij}$ is assigned a weight $W(F,\dm,\bm\Theta, \bm\lambda)$, where $\bm\Theta$ is the parameter set that specifically includes the intrinsic pulse width $\omega$ and the scattering broadening timescale $\tau$ (i.e.,\(\bm\Theta=\{\omega, \tau\}\))}, and the parameter set $\bm\lambda$ characterizes the considered fluence and DM distribution, where its specific parameter composition varies depending on the model, with detailed descriptions provided in Section~\ref{sec:fdm_model}. Essentially, the independence of $\tau$, $\omega$ from $F$, and DM are assumed in the injected data-sets \citep{catalog1, Cui+2025ApJ}. While this is approximately true for the full set of 5 million simulated bursts, we acknowledge that CHIME imposed certain cuts  during the injection campaign, which may introduce weak correlations between parameters in the detected subset. Nevertheless, these effects are expected to be minor and do not significantly alter our results. The selection effect can be described by a selection function.  
From a physical perspective, the selection function should be a continuous function. Therefore, in the phase space formed by $\snr$ and $\dm$, for any point $P$, there exists a sufficiently small neighborhood such that the variation in the value of the selection function is adequately minimal. 
Thus in the neighbourhood of point $P$, the theoretical values predicted by different models (after the action of the selection function) will be proportional only to the event rate $R$.
Using this concept, the theoretical value of the second model can be computed within the vicinity of point $P$. By selecting a neighborhood where each data point corresponds to only one FRB from the injection model, the number of FRBs observed from any designed model in the $(i,j)$ region is given by:
\begin{eqnarray} 
\label{eqn:weight} 
\zeta_{ij} = \sum^{n_{ij}}_{m=1} W_\mathrm{m}(F, \dm, \bm\Theta, \bm\lambda),
\end{eqnarray}
This quantity represents a prediction of the observed $\snr$-$\dm$ distribution in Catalog~1. Each injected burst $m$ is associated with its "intrinsic" $F$ and $\dm$ values, and the bursts also have "observed" $\snr$ values. Importantly, only the subset of injected bursts that were detected should be included in this summation, as these are the events that contribute to the observed distributions. As such, each weighted burst is assigned to a bin in $\zeta_{ij}$, which is determined by its S/N and DM, and its weight depends on the parameters of the given models. This allows for a direct comparison with ${n}_{ij}$ in the likelihood function (see Eq.~(\ref{eqn:Poisson_likelihood})).
Different candidate population models $\zeta_{ij}$ can be evaluated by this likelihood.

Specifically, these weights are calculated as the ratio of the number of injected bursts to the total number of bursts that would be observed in the sky during the survey, based on a given set of model parameters $\bm\lambda$. Therefore, the number of bursts to be injected for each synthetic burst with properties $F$ and $\dm$ {should match $W(F,\dm,\bm\Theta,\bm\lambda)$ in order to accurately reproduce the observed sky population. }{Ultimately,  for each synthetic burst $m$ with properties $F$ and $\dm$, $W(F,\dm,\bm\Theta,\bm\lambda)$
is calculated by dividing the total number of detected injected bursts by the total initial population of injected bursts,}
\begin{eqnarray} 
	\label{eqn:weight_m} 
	W(F,\dm,\bm\Theta,\bm\lambda) = \frac{R(F,\dm,\bm\Theta |\bm\lambda) \Delta t}{R_\mathrm{inj}(F,\dm,\bm\Theta)}.
\end{eqnarray}
Here, $R(F,\dm,\bm\Theta|\bm\lambda)$ represents the burst rate per year, which is multiplied by the survey duration ($\Delta t = 214.8$ days) to obtain the total number of observed bursts during the survey. On the other hand, $R_\mathrm{inj}(F,\dm,\bm\Theta)$ corresponds to the injected population of bursts. This is obtained by multiplying the normalized  population distribution of injected synthetic bursts, $P_{\text{inj}}(F, \mathrm{DM}, \bm\Theta)$, by the total number of injected FRBs ($N_{\rm inj}$), while also correcting for the injection efficiency ($\epsilon_{\rm inj}$) and the fraction of the sky observed ($f_{\rm sky}$). This weighting is applied to all injected bursts, regardless of whether they were detected or not.  
Thus, the final expression for the weight of each burst is
\begin{eqnarray} 
	\label{eqn:weight} 
	W(F,\dm,\bm\Theta,\bm\lambda) = \frac{\Delta t f_\mathrm{sky}}{\epsilon_\mathrm{inj} N_\mathrm{inj} } \frac{R(F,\dm|\bm\lambda)}{P_\mathrm{inj}(F)P_\mathrm{inj}(\dm)} \frac{P_\mathrm{fid}(\bm\Theta)}{P_\mathrm{inj}(\bm\Theta)},\nonumber\\
\end{eqnarray}
where $P_{\text {inj }}(F, \mathrm{DM}, \bm\Theta)$ has been factorized  into $P_\mathrm{inj}(F) P_\mathrm{inj}(\dm) P_\mathrm{inj}(\bm\Theta) $.   Both the injection model  and the fiducial model $P(\bm\Theta)$  are described by lognormal distributions \citep{catalog1},
\begin{eqnarray} 
	\label{eqn:inj} 
	P(\boldsymbol{\Theta}) = \frac{1}{\sigma(\boldsymbol{\Theta}/\eta)\sqrt{2\pi}} \exp\left[-\frac{\ln^2(\boldsymbol{\Theta}/\eta)}{2\sigma^2}\right],
\end{eqnarray}
where $\eta$ and $\sigma$ represent the mean and standard deviation of the distribution.
This distribution  is normalized such that $\int \mathrm{d}\bm\Theta P(\bm\Theta) = 1$.  
Similarly, the injected distribution of $\dm$, $P_\mathrm{inj}(\dm)$, is also described by Eq.~(\ref{eqn:inj}). The parameters used for both injected and fiducial distributions are listed in Table \ref{tab:param_1}.
For fluence $F$, the injected distribution follows  a simple power-law, $P_\mathrm{inj}(F) \propto (F/F_0)^{-2}$, where  $F_0 = 0.2$ Jy $\cdot$ ms \citep{catalog1,merryfield+2022}.
\begin{deluxetable}{cccc}  
	\tablecaption{ The parameters of the injected  and fiducial distributions. }  
	\label{tab:param_1}  
	\tablehead{\colhead{Parameter} & \colhead{$\eta$} & \colhead{$\sigma$}&\colhead{Reference}}   
	\startdata  
	\multicolumn{4}{c}{\textbf{Injection model}} \\[0.6em]  
	$\dm$ & 462 pc cm$^{-3}$ & 0.76&[1]\\[0.5em]  
	$\omega$ & 1.06 ms& 1.04&[1]\\[0.5em]  
	$\tau$ & 0.74 ms &2.06&[1]\\[0.6em]  
	\multicolumn{4}{c}{\textbf{Fiducial model}} \\[0.6em]  
	$\omega$ & 1.00 ms& 0.97&[2]\\[0.5em]  
	$\tau$ & 2.02 ms &1.72&[2]\\[0.5em]  
	\enddata  
	\tablenotetext{ }{Note. {The references are as follows: [1] \citet{merryfield+2022}, [2] \citet{catalog1}.} }  
\end{deluxetable}

Finally, the joint rate distribution of fluence $F$ and $\dm$, $R(F,\dm|\bm\lambda)$ in Eq.(\ref{eqn:weight}), is defined as 
\begin{eqnarray}
	\label{eqn:rate_fdm_distr}
	R(F,\dm|\bm\lambda) = \int \mathrm{d}z\, R(F,z|\bm\lambda_1) P(\dm|z,\bm\lambda_2).
\end{eqnarray}
Within the integrand, $R(F,z|\bm\lambda_1)$ represents the joint rate distribution of fluence and redshift, while $P(\dm|z,\bm\lambda_2)$ is the probability distribution of $\dm$ given the redshift. The parameters \(\bm{\lambda_1}\) and \(\bm{\lambda_2}\) encapsulate different aspects of the model. Specifically, \(\bm{\lambda_1}\) refers to the set of parameters that describe the fluence and redshift relationship, which  govern the energy distribution, the volume element, the FRB rate per comoving volume, and other cosmological factors affecting the fluence-redshift relationship. 

There exists a direct conversion between the FRB's energy and its fluence,
\begin{eqnarray} 
	\label{eqn:energy_function}
	E=\frac{4\pi D^2_L(z)}{(1+z)^{2+\alpha}}\Delta\nu F.
\end{eqnarray}
where $D_L(z)$ is the luminosity distance, $\alpha$ is the spectral index\footnote{Here, we introduce the spectral index $\alpha$ in our model to capture realistic variations in FRB spectra ,by following  \cite{shin+2023}. However, this parameters is not present in the injected datasets and thus are not explicitly included in the completeness function. We assume that spectral behavior primarily affects detectability through the effective fluence at CHIME band and is captured indirectly via the fluence–DM–S/N reweighting. A more rigorous treatment would require an updated injection dataset with varying spectral parameters, which is beyond the scope of this work.}, and $\Delta\nu$ is the observer-frame frequency bandwidth, which we take to be $\Delta\nu = 1$ GHz. The joint rate distribution of fluence and redshift is given by \citep{shin+2023}
\begin{eqnarray}
	\label{eqn:R(F,z)}
	R(F, z|\bm\lambda_1) 
	= P(E) \chi(z)^2 \frac{\mathrm{d}\chi}{\mathrm{d}z} \Phi(z) 
	\frac{4 \pi D_L^2(z)}{(1+z)^{2+\alpha}} \Delta\nu,
\end{eqnarray}
where $P(E)$ is the energy distribution, $\chi(z)$ is the comoving distance, and $\chi(z)^2 \frac{\mathrm{d}\chi}{\mathrm{d}z}$ represents the volume element $\frac{\mathrm{d}V}{\mathrm{d}z}$. $\Phi(z)$ is the FRB rate per comoving volume (volumetric event rate).
The probability distribution of $\dm$ given redshift $z$ is  given by \citep{shin+2023}
\begin{equation}
	\label{eqn:dm}
	P(\dm|z,\bm\lambda_2) = P(\dm = \dm_\mathrm{EG} + \overline{\dm}_\mathrm{MW} | z, \bm\lambda_2),
\end{equation}
where \(\bm{\lambda_2}\) describes the distribution of $\dm$, $\dm_\mathrm{EG}$ represents contributions from extragalactic sources conditional on the redshift $z$, and $\overline{\dm}_\mathrm{MW}$ is the average contribution from the Milky Way.

The above formulation establishes the foundation for modeling the fluence and dispersion measure distributions of FRBs. In the next section, we will describe the specific models for the FRB energy distribution, redshift evolution, and dispersion measure components in detail. These models are essential for interpreting the observed FRB population and extracting their intrinsic properties.

\section{The models \label{sec:fdm_model}}

\subsection{Energy Distributions \label{subsec:E_distr}} 
In this work, we explore two different energy distribution models for FRBs. The first one is Schechter function, which follows a power law with a high-energy exponential cut off \citep{schechter1976},
\begin{eqnarray}
\label{eqn:schechter_function}
	P(E)\propto\left(\frac{E}{E_\mathrm{c}}\right)^{\gamma}\exp({-\frac{E}{E_\mathrm{c}}}),
\end{eqnarray}
where $\gamma$ is the power-law index, and $E_\mathrm{c}$ is the characteristic cutoff energy. And because of this, in the following text, we refer to it as the cutoff power-law function. This model describes a scenario in which lower-energy FRBs follow a power-law distribution, while high-energy FRBs become increasingly rare due to an exponential suppression. The cutoff power-law function has been widely used in astrophysical studies to describe luminosity functions of galaxies, gamma-ray bursts, and other transient phenomena, making it a natural choice for modeling FRB energy distributions.

While most previous studies have focused on the cutoff power-law function, we also consider the broken power law function as a more flexible alternative, which has been widely used in gamma-ray bursts tudies \citep{2016ApJ...820...66D} and may also be applicable to FRBs.  The smooth  broken power law is expressed as
\begin{eqnarray} 
\label{eqn:broken_power_law_function}
	P(E)\propto\left[\left(\frac{E}{E_b}\right)^{\omega \gamma_1}+\left(\frac{E}{E_b}\right)^{\omega\gamma_2}\right]^{1/\omega},
\end{eqnarray}
where $\gamma_1$ and $\gamma_2$ are the power-law indices before and after the break energy $E_b$, and $\omega$ defines the sharpness of the break, we fix $\omega=3$. Similarly, here we assume that $P(E)$ does not evolve with redshift.

Both above models initially assume that the energy distribution does not evolve with redshift. However, some studies suggest that FRB energies may evolve over  redshift  \citep{chen+2024_redshift,Zhang+2025}. To account for this possibility, we also consider a scenario where the FRB energy evolve over  redshift as $E\propto(1+z)^k$, where $k$ is an evolution parameter to be determined. When $k = 0$, the energy distribution remains unchanged with redshift, corresponding to the standard non-evolving case. When $k \neq 0$, the energy distribution is redshift-dependent, implying possible cosmic evolution of FRB properties. By fitting both cases, we can assess whether there is significant evidence supporting redshift evolution in the FRB energy distribution.

\subsection{Redshift Distribution Models} \label{subsec:SFR_model}
The formation history of FRBs remains uncertain, but it is reasonable to expect that their volumetric event rate $\Phi(z)$ is linked to the cosmic star formation rate history $\sfr(z)$, as many FRB progenitor models suggest young magnetars as their likely origin. However, alternative scenarios, such as FRBs originating from evolved stellar populations, cannot be ruled out. To explore the potential redshift evolution of FRB populations, we consider five different models for $\Phi(z)$, each incorporating a different relationship between FRB formation and cosmic star formation history.  

The simplest assumption is that the FRB formation rate directly follows the cosmic star formation history, meaning FRBs are predominantly produced by young stellar populations. This scenario aligns with the hypothesis that FRBs originate from young magnetars, which are expected to form shortly after the deaths of massive stars. In this model, the volumetric event rate follows
\begin{equation} \label{eqn:Phi(z)_2} 
\Phi(z) = \frac{\Phi_0}{1+z}\frac{\sfr(z)}{\sfr(0)}. 
\end{equation}
where $\Phi_0$ is the FRB volumetric rate at $z=0$, and $\sfr(z)$ is the cosmic star formation rate \citep{madaudickinson2014}:
\begin{equation} \label{eqn:SFR(z)}
 \sfr(z) = 1.0025738 \frac{(1+z)^{2.7}}{1+\left(\frac{1+z}{2.9}\right)^{5.6}}.
 \end{equation}
This model provides a baseline scenario where FRBs perfectly track the SFH, without additional evolutionary effects.

A more flexible modification to the direct SFH model is to allow the FRB formation rate to scale with a power-law index. This model is had been used in FRB population studies by \cite{james+2022_energy,shin+2023}, the latter being the foundation of our analysis methodology. The volumetric event rate in this case is given by
\begin{equation}
\label{eqn:phi(z)_1a}
\Phi(z) = \frac{\Phi_0}{1+z} \left( \frac{\sfr(z)}{\sfr(0)} \right)^n,
\end{equation}
where $n$ is a free parameter controlling how strongly the FRB rate correlates with the SFH. When $n = 1$, the FRB formation rate directly tracks the SFH, recovering the pure SFH model.  When $n > 1$, FRBs exhibit a stronger-than-SFH evolution, meaning their occurrence increases more rapidly at higher redshifts. When $n < 1$, FRBs evolve more slowly compared to the SFH, suggesting that additional astrophysical effects may be influencing their redshift distribution. This model provides a natural extension of the direct SFH model while maintaining a straightforward scaling relationship between FRB formation and stellar birth rates.

An alternative approach, frequently considered in gamma-ray burst studies, introduces an additional $(1+z)$ evolutionary factor to allow for stronger redshift dependence:
\begin{equation}
\label{eqn:phi(z)_1b}
\Phi(z) = \frac{\Phi_0}{1+z} \frac{\sfr(z)}{\sfr(0)} (1+z)^{\beta},
\end{equation}
where $\beta$ is a free parameter governing additional redshift evolution. $\beta = 0$ reduces to the pure SFH-tracking model. When $\beta > 0$, FRBs exhibit an enhanced occurrence rate at higher redshifts beyond what is expected from SFH alone. When $\beta < 0$, FRBs become less frequent at higher redshifts, potentially indicating an environmental dependence such as metallicity constraints or evolving magnetar formation pathways. By fitting both $n$ and $\beta$, we can determine whether FRB formation exhibits a stronger or weaker correlation with the cosmic SFH than previously assumed.

While some FRBs may originate from young magnetars, observational evidence suggests that a subset of FRBs might be associated with older stellar populations, as indicated by their presence in globular clusters, such as FRB 20200120E in the M81 galaxy \citep{bhardwaj+2021_m81, kirsten+2022_m81}. If some FRBs arise from compact object mergers or other long-lived progenitors, their formation would exhibit a time delay relative to the cosmic SFH. To model this, we assume that the FRB event rate follows a delayed star formation history, and the volumetric event rate in this case is given by
\begin{equation}
\label{eqn:phi(z)_delay}
\Phi(z) = \frac{\Phi_0}{1+z}\frac{\sfr_\mathrm{d}(z)}{\sfr_\mathrm{d}(0)},
\end{equation}
where the $\sfr_\mathrm{d}(z)$ is a convolution of the SFH and a delay time distribution $f(\tau)$:
\begin{equation} \label{eqn:delay_model} 
	\sfr_\mathrm{d}(z) \propto \int_z^\infty \sfr(z') f(t(z') - t(z)) \frac{\mathrm{d}t}{\mathrm{d}z'} \mathrm{d}z', 
\end{equation}
where $t(z)$ is the cosmic time corresponding to redshift $z$, and $f(\tau)$ describes the probability distribution of delay times. We consider two different delay distribution:

1. Log-normal delay model:
\begin{equation} \label{eqn:Gaussian_delay} 
	f(\tau) = \frac{1}{\tau \sigma_{\tau} \sqrt{2\pi}} \exp\left(-\frac{(\mathrm{ln}\tau - \mathrm{ln}\bar{\tau})^2}{2\sigma_{\tau}^2}\right), 
\end{equation}
where $\bar{\tau}$ is the characteristic delay time, and $\sigma_{\tau}$ controls the spread of the delay distribution. 

2. Power-law delay model:
\begin{equation} \label{eqn:Gaussian_delay} 
	f(\tau) \propto \tau^{\alpha_\tau}
\end{equation}
where $\alpha_\tau$ is a power-law time delay index. To reduce the computational load during the fitting process, we set the minimum time delay $\tau_\mathrm{min}$ to a fixed value of 30 Myr. We have tried several different values and found that the results do not differ significantly.

For a given redshift $z$, the lookback time is
\begin{equation}
\label{eqn:lookback_time}
t=\int_0^z\frac{t_H}{(1+z)\sqrt{\Omega_m(1+z)^3+ \Omega_\wedge}}\mathrm{d}z,
\end{equation}
where $t_H=1/H_0$ is the Hubble time, the Hubble constant is taken as $H_0=67.4$ km s$^{-1}$ Mpc$^{-1}$. In a flat $\wedge$ CDM universe, matter density $\Omega_m=0.315$ and cosmological constant $\Omega_\wedge=0.685$ \citep{planck2018vi}.
This model captures scenarios where FRBs arise from evolved stellar systems, such as binary mergers, old magnetars, or white dwarf accretion events.

Finally, we consider a more general power-law evolution model, where the FRB volumetric event rate follows a simple power-law dependence on redshift, independent of the cosmic SFH. This model had been considered by \cite{Lu+2019} and aligns with the suggestions from recent non-parametric analyses \cite{2019JHEAp..23....1D,chen+2024_redshift,Zhang+2025}, which indicate  that the redshift evolution of $\Phi(z)$ may follow a simple relationship proportional to $(1+z)^\delta$.
The volumetric event rate in this case is parameterized as
\begin{equation} \label{eqn:Phi(z)_3} 
	\Phi(z) = \frac{\Phi_0}{1+z}(1+z)^\delta, 
\end{equation}
where $\delta$ is a free evolution parameter governing how the FRB population changes with cosmic time. 
This model serves as a non-SFH-dependent alternative that allows us to directly probe whether FRBs exhibit a distinct redshift evolution trend independent of massive star formation. By comparing this model to the SFH-based models, we can assess whether FRBs are more tightly coupled to stellar evolution or if their occurrence follows a more universal cosmological trend.

\subsection{Dispersion Measure Distributions \label{subsec:dm_dist}} 
The redshift of an FRB is crucial for understanding its population and cosmological evolution. However, the majority of FRBs lack direct redshift measurements. One solution to this problem is to use the DM as a proxy for redshift. The observed DM, which reflects the total column density of free electrons along the line of sight, consists of contributions from several components: Milky Way contributions (from the disk, halo, and other structures) and Extragalactic contributions (from the cosmic ionized gas and the FRB's host galaxy).
The total observed DM can be expressed as
\begin{eqnarray}
\label{eqn:ob_dm_model}
	\dm=\dm_\mathrm{cosmic}+\overline\dm_\mathrm{MW}+\dm_\mathrm{host},
\end{eqnarray}
where $\overline\dm_\mathrm{MW}$, the average contribution from the Milky Way, combines the disk and halo contributions. In our model, we fix $\overline\dm_\mathrm{MW}$ at 80 pc cm$^{-3}$ by following \cite{shin+2023}\footnote{The approach of fixing $\mathrm{DM}_{\rm MW} = 80 \, \mathrm{pc \, cm^{-3}}$ has been adopted from \cite{shin+2023}, who acknowledged that while this assumption introduces some uncertainty, the impact on the model fitting is secondary to statistical uncertainty. In our study, we also use this baseline value of $\mathrm{DM}_{\rm MW}$, and we note that the uncertainty associated with it is absorbed by the fitted $\mathrm{DM}_{\rm host}$, the host galaxy's dispersion measure. This approach ensures that the uncertainty in $\mathrm{DM}_{\rm MW}$ does not significantly affect the final model parameters, as the host galaxy dispersion measure compensates for this uncertainty.}. For the extragalactic (EG) contributions, we define $\dm_\mathrm{EG} = \dm_\mathrm{cosmic} + \dm_\mathrm{host}$, where $\dm_\mathrm{cosmic}$ accounts for the ionized gas in the intergalactic medium, and $\dm_\mathrm{host}$ is the contribution from the FRB’s host galaxy.

Define a $\Delta_\dm$, which is the deviation from the mean, such that
\begin{equation} 
\label{eqn:dm_cosmic}
	\dm_{\mathrm{cosmic}}=\Delta_\dm \left<\dm_{\mathrm{cosmic}}\right>(z),
\end{equation}
where the expectation value of the $\left<\dm_{\mathrm{cosmic}}\right>(z)$ is derived by performing calculations on the mean density of ions $\bar{n}_e$ and related cosmological parameters, which is given by \citep[see e.g.,][]{ioka+2003_dmz,inoue+2004_dmz,Zheng+2014}
\begin{equation}
\label{eqn:dm_cosmic_1}
	\left<\dm_{\mathrm{cosmic}}\right>(z)= \int_0^z\frac{c \bar{n}_e(z') \mathrm{d}z'}{H_0 (1+z')^2 \sqrt{\Omega_m (1+z')^3 + \Omega_\Lambda}}.
\end{equation}
Using this, one can obtain the distribution $P(\dm_\mathrm{cosmic}|z)$ for the cosmic DM component at a given redshift,
\begin{equation} 
\label{eqn:dmcosmic_distr}
	P(\dm_\mathrm{cosmic}|z)=P(\varDelta_\mathrm{DM})\frac{1}{\left<\dm_{\mathrm{cosmic}}\right>(z)}.
\end{equation}
The probability of deviations from the mean, $P(\Delta_\dm )$, then, is given by
\begin{equation}
\label{eqn:mean_dm}
	P(\Delta_\dm|z)=\kappa \Delta_\dm ^{-B}\exp\left[-\frac{( \Delta_\dm^{-A}-C_0)^2}{2A^2\sigma_\dm^2}\right],
\end{equation}
where $A = 3$, $B = 3$, and $C_0$ is tuned such that the expectation value of the distribution is unity. The effective standard deviation $\sigma_\dm$ is defined as $\sigma_\dm = Fz^{-0.5}$, with $F = 0.32$ based on the results of \citet{macquart+2020}.

The contribution from the host galaxy's DM, $\dm_\mathrm{host}$, carries valuable information about the interstellar plasma and the environment around the FRB. Although the exact distribution of the host galaxy's DM is not directly observed, it is commonly assumed that $\dm_\mathrm{host}$ follows a lognormal distribution \citep{macquart+2020}.
The probability distribution for $\dm_\mathrm{host}$ is expressed as
\begin{equation} 
\label{eqn:P_DMhost}
	P(\dm_{\mathrm{host}}')=\frac{1}{\dm_{\mathrm{host}}'}\frac{1}{\sigma_{\mathrm{host}}\sqrt{2\pi}}
	e^{-\frac{(\log \dm_{\mathrm{host}}'-\mu_{\mathrm{host}})^2}{2\sigma^2_{\mathrm{host}}}},
\end{equation}
where $\mu_{\mathrm{host}}$ and $\sigma_{\mathrm{host}}$ are the mean and standard deviation of the lognormal distribution, respectively. 
To account for the redshift effect,  the correction of $\dm_\mathrm{host}=\dm_\mathrm{host}'/(1+z)$ is applied.

Since we use a fixed value for $\overline\dm_\mathrm{MW}$, and given the relation $\dm = \dm_\mathrm{cosmic} + \overline\dm_\mathrm{MW} + \dm_\mathrm{host}$, to compute $P(\dm|z, \bm\lambda_2)$ as used in Eq.(\ref{eqn:dm}), it is sufficient to calculate $P(\dm_\mathrm{EG}|z)$. The extragalactic contribution $P(\dm_\mathrm{EG}|z)$ is determined by the convolution of $P(\dm_\mathrm{cosmic}|z)$ and $P(\dm_\mathrm{host}|z)$ as follows
\begin{eqnarray}
\label{eqn:dm_EG}
	P(\dm_{\mathrm{EG}}|z)
	&=& \int \mathrm{d}\dm_{\mathrm{cosmic}}P(\dm_{\mathrm{host}}=\dm_\mathrm{EG} \nonumber\\
	&-& \dm_\mathrm{cosmic}|z)
	P(\dm_{\mathrm{cosmic}}|z).
\end{eqnarray}

In this section, we have introduced and discussed the key models used to describe the energy distributions, redshift distributions, and DM distributions of FRBs. These models form the foundation for understanding the population and cosmological evolution of FRBs. We explored two energy distribution models—the cutoff power-law function and the broken power law—and considered the possibility of redshift evolution in FRB energies. Additionally, we presented five redshift distribution models, ranging from direct tracking of the cosmic star formation rate history to more complex scenarios involving delayed star formation and power-law evolution. Finally, we detailed the DM distribution models, which account for contributions from the Milky Way, the intergalactic medium, and the host galaxy, providing a crucial link between observed DMs and FRB redshifts.
These models collectively enable a comprehensive analysis of FRB populations, allowing us to probe their origins, evolution, and environmental dependencies. In the next section, we will present the results of fitting these models to observational data, shedding light on the underlying properties and behaviors of FRBs across cosmic time.

\begin{figure*}
	\begin{center}
		\centering 
		\includegraphics[width=1\textwidth]{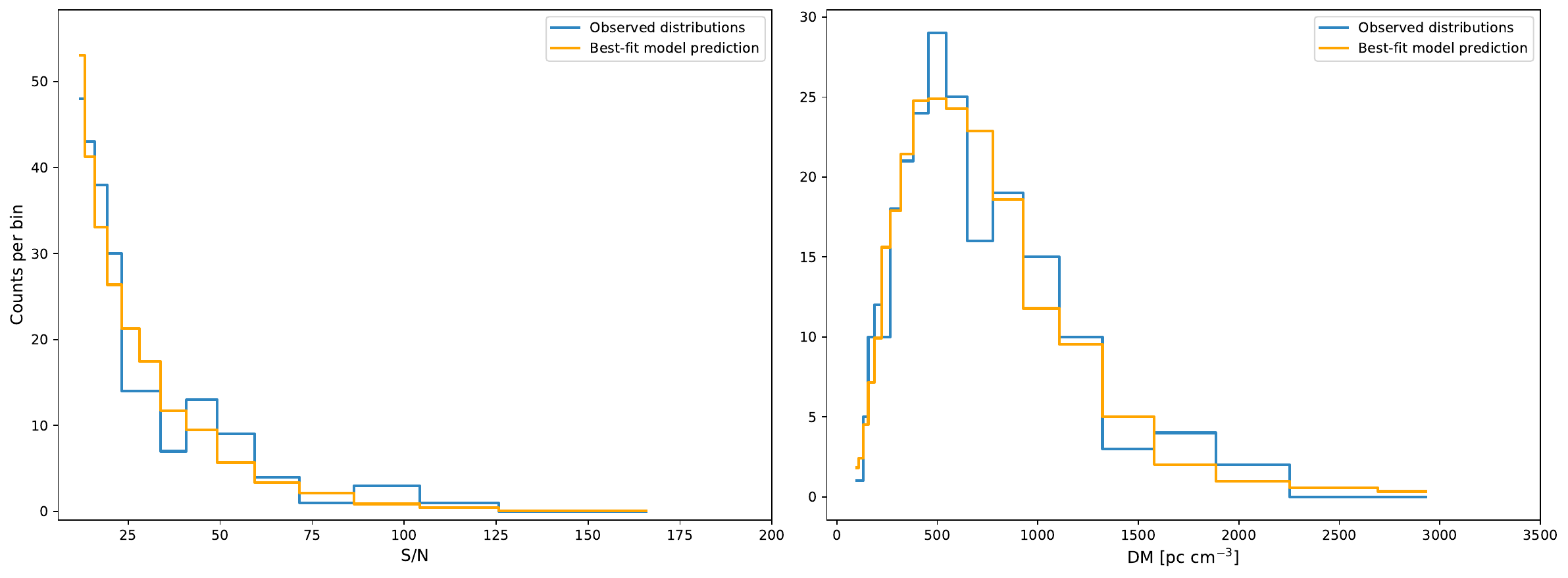}\\
		\caption{Observed distributions (blue lines) compared with the best-fit model predictions (orange lines). Left panel: the S/N distribution is shown for the DM samples; right panel: the DM distribution is shown for the S/N samples. The best-fit model predictions appears to be consistent with the observed data.
			\label{fig:1}}
	\end{center}
\end{figure*}
\section{RESULTS \label{sec:RESULT}} 
In this section, we present the detailed results of our analysis, focusing on the energy distribution and redshift distribution models for FRBs introduced in Section~\ref{sec:fdm_model}. 
We consider two models for energy distributions—cutoff power-law function and broken power-law function—along with five different models for the redshift idistribution. 
Our goal is to understand the population and cosmological evolution of FRBs by fitting these models to observational data. We begin by assuming that the energy distribution \( P(E) \) does not evolve with redshift. 
However, we also explore the potential for redshift evolution in the energy distribution, using a paramete \( k \) 
 to model energy evolution as  \( E \propto (1+z)^k \).

\begin{figure*}
	\begin{center}
		\centering 
		\includegraphics[width=0.5\textwidth]{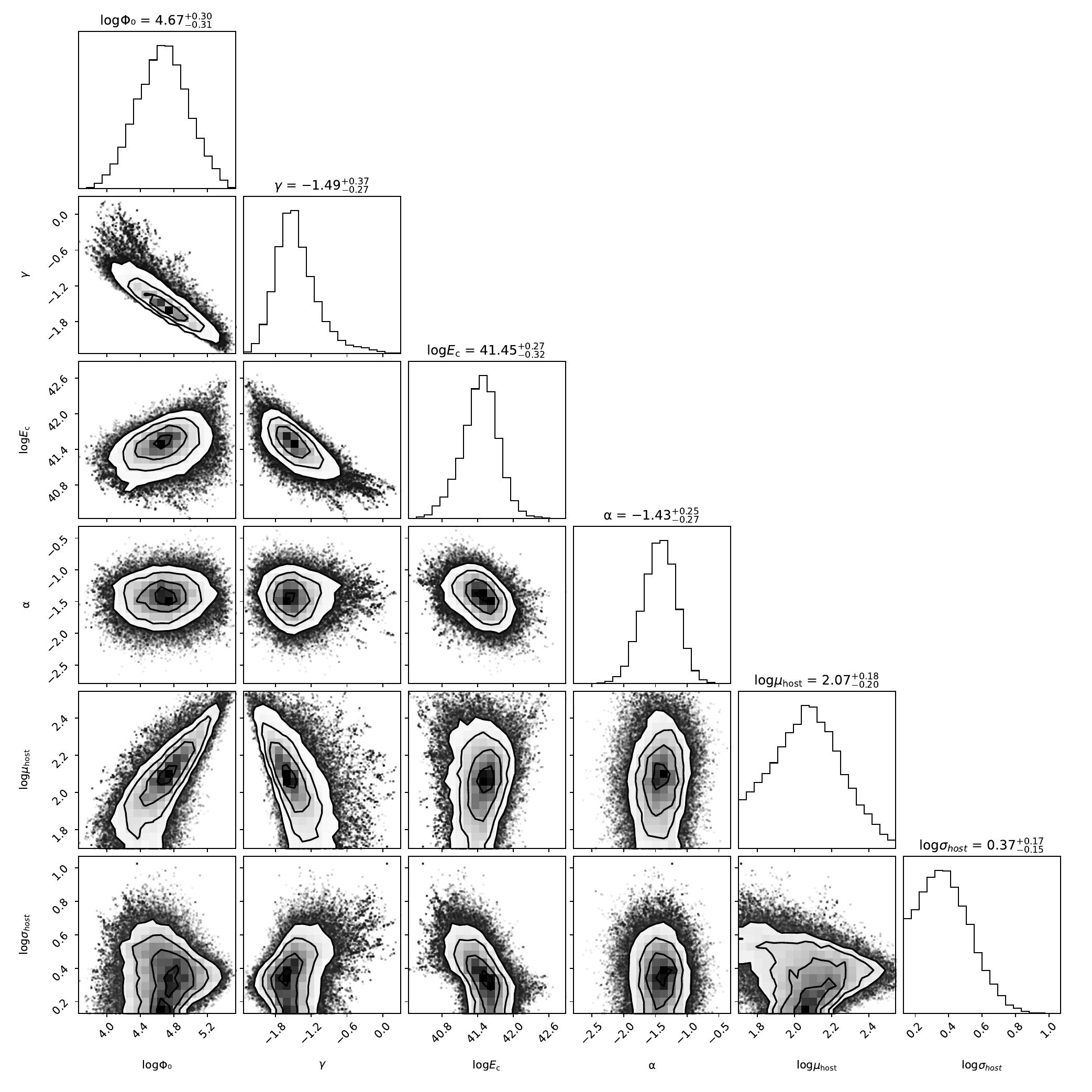}
		\includegraphics[width=0.49\textwidth]{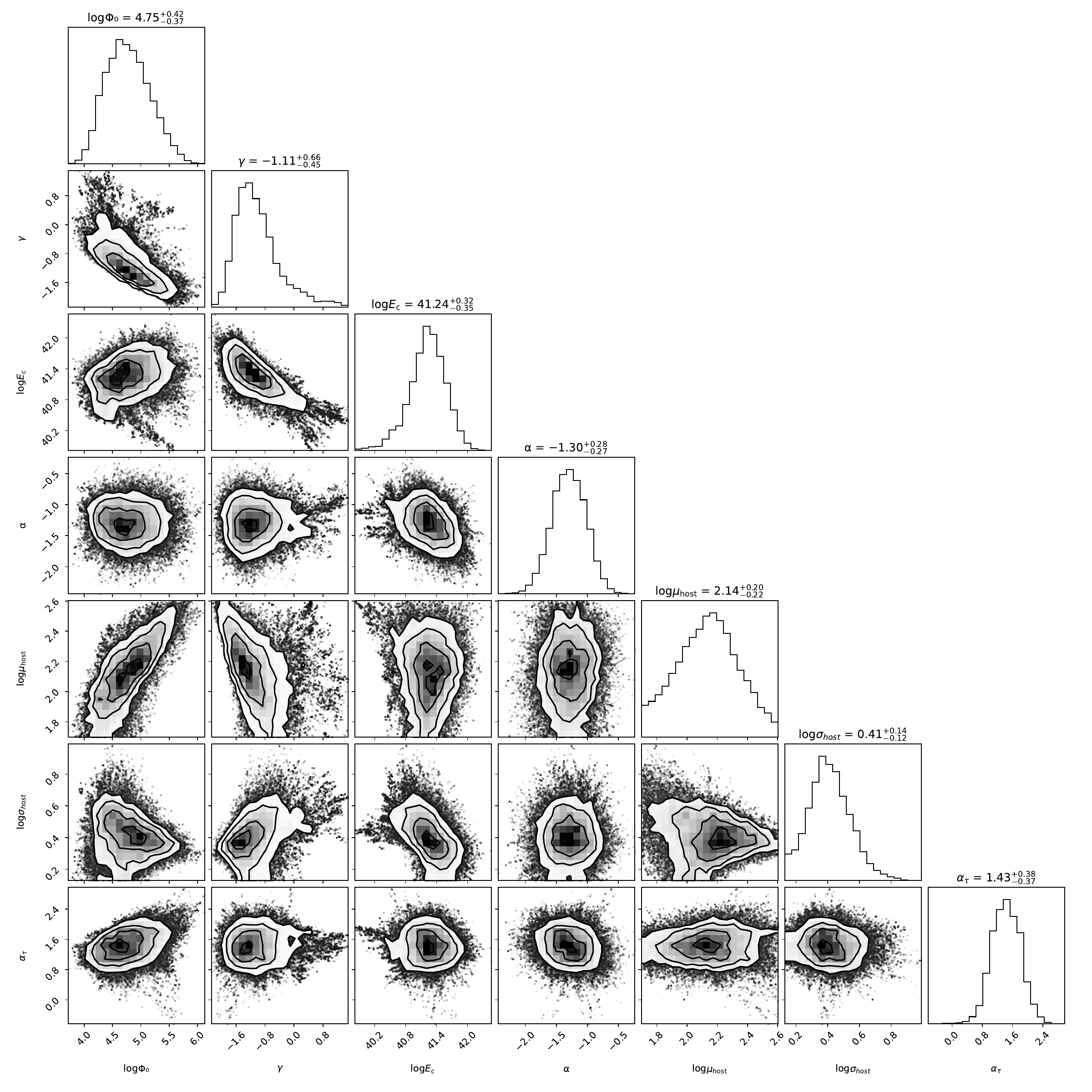}\\
		\caption{Corner plot of the results of the MCMC run for Schechter function. Left panel: redshift model $\Phi(z)\propto \sfr(z)$; right panel: power-law delay model.
			\label{fig:MCMC1}}
	\end{center}
\end{figure*}

\begin{figure*}
	\begin{center}
		\centering 
		\includegraphics[width=0.49\textwidth]{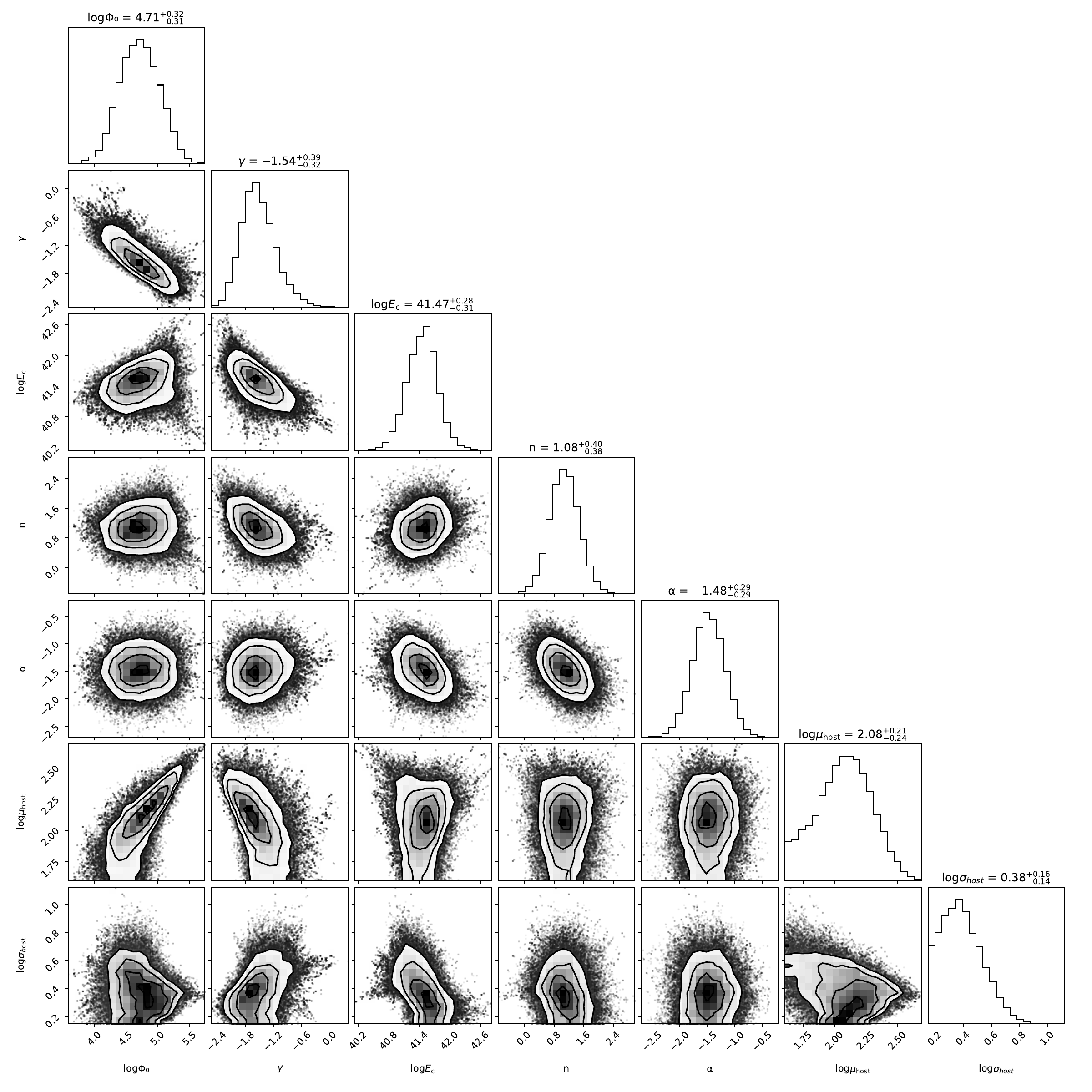}
		\includegraphics[width=0.49\textwidth]{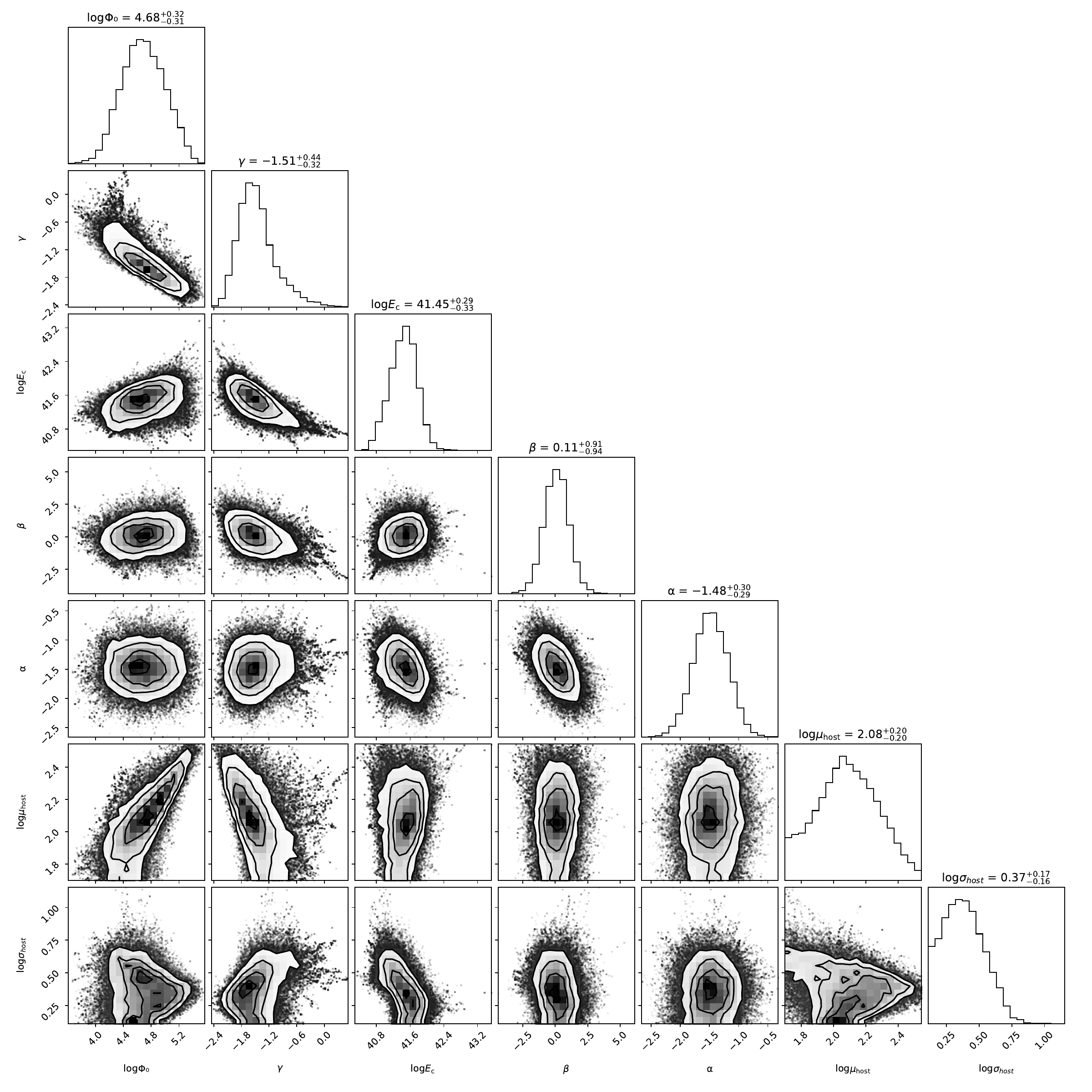}\\
		\caption{Corner plot of the results of the MCMC run for Schechter function. Left panel: redshift model $\Phi(z)\propto [\sfr(z)]^n$; right panel: redshift model $\Phi(z)\propto \sfr(z)(1+z)^{\beta}$.
			\label{fig:MCMC2}}
	\end{center}
\end{figure*}

\begin{figure*}
	\begin{center}
		\centering 
		\includegraphics[width=0.49\textwidth]{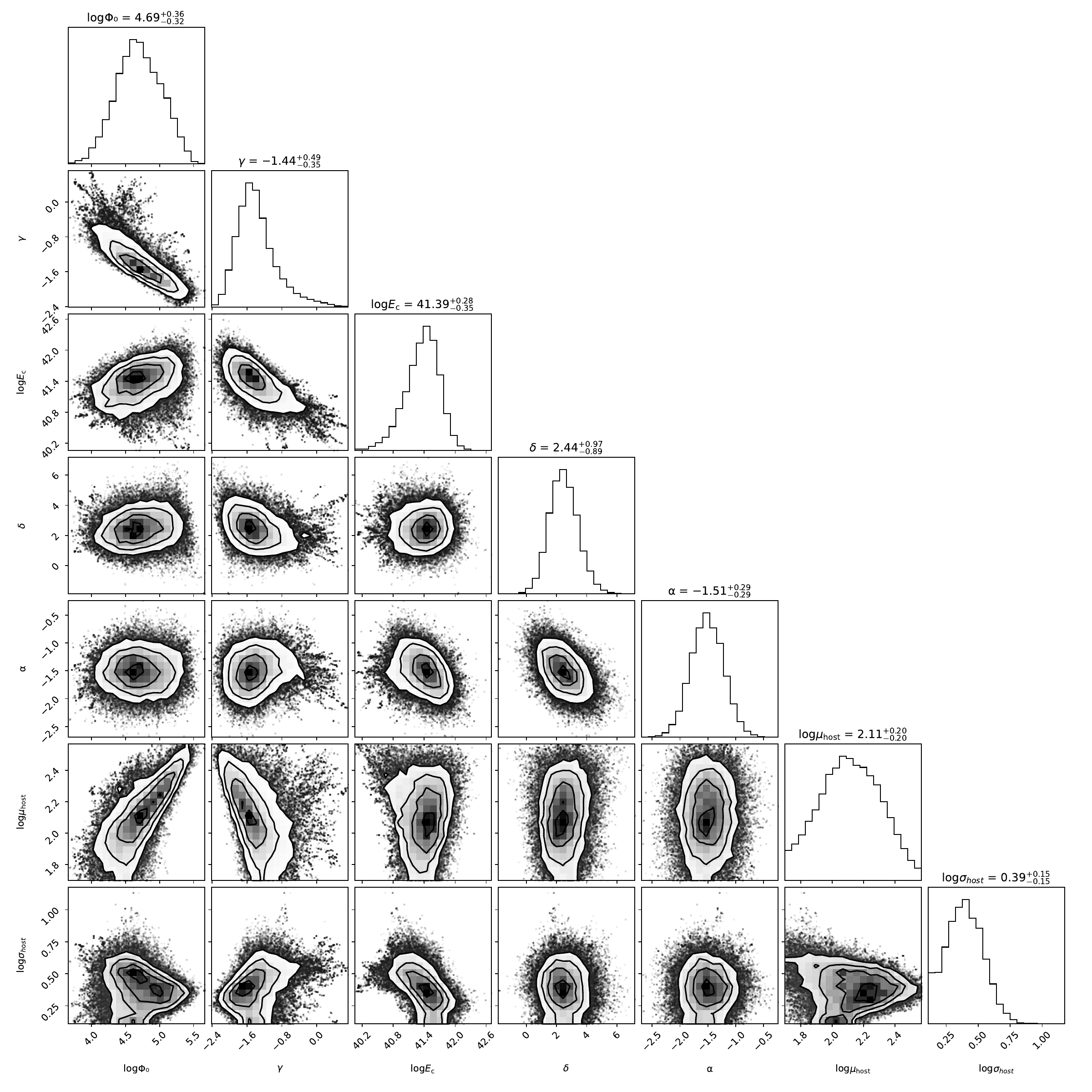}
		\includegraphics[width=0.49\textwidth]{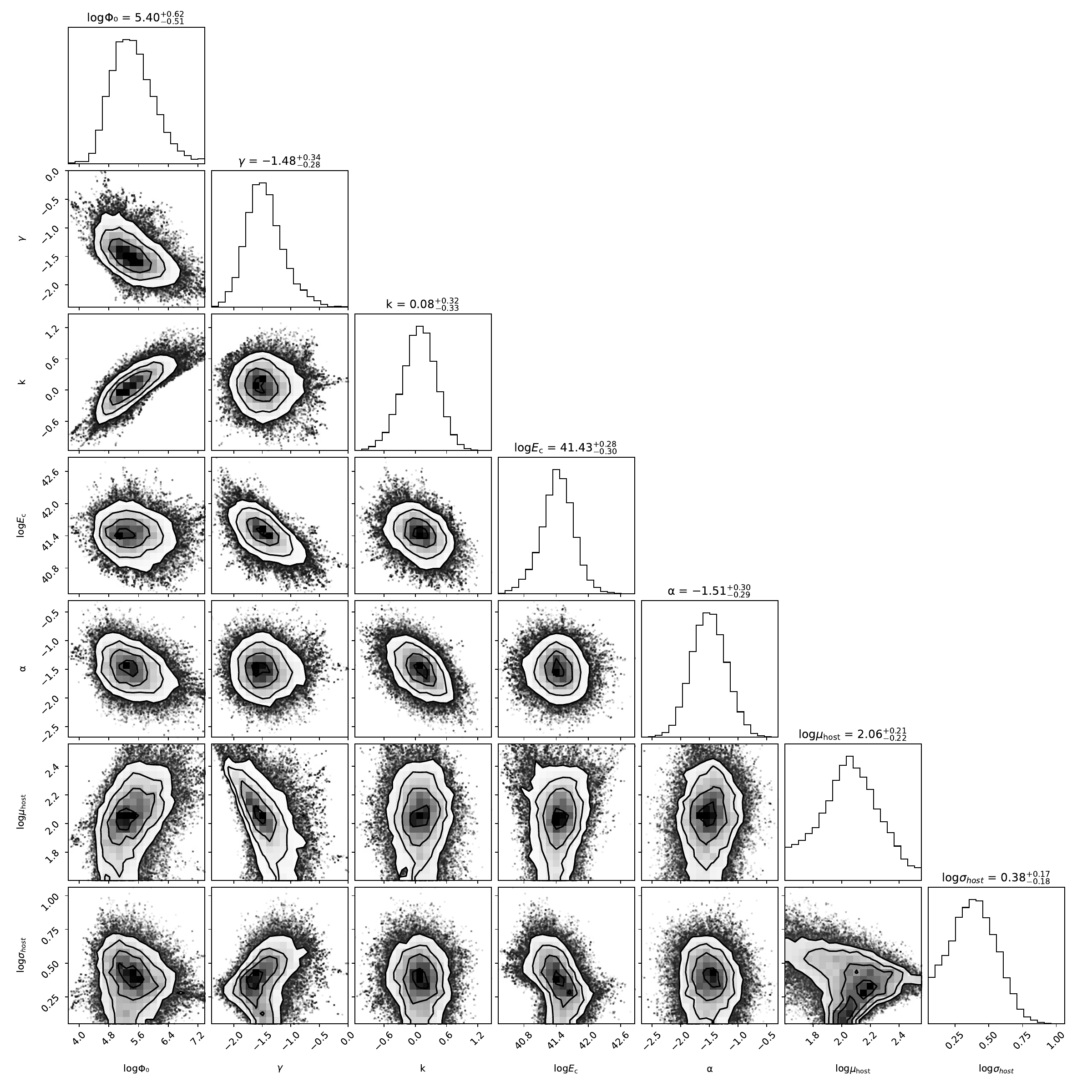}\\
		\caption{Corner plot of the results of the MCMC run for Schechter function. Left panel: redshift model $\Phi(z)\propto (1+z)^\delta$; right panel: energy evolve with redshift ($E\propto(1+z)^k$ ), redshift model $\Phi(z)\propto \sfr(z)$.
			\label{fig:MCMC3}}
	\end{center}
\end{figure*}

\begin{figure*}
	\begin{center}
		\centering 
		\includegraphics[width=0.49\textwidth]{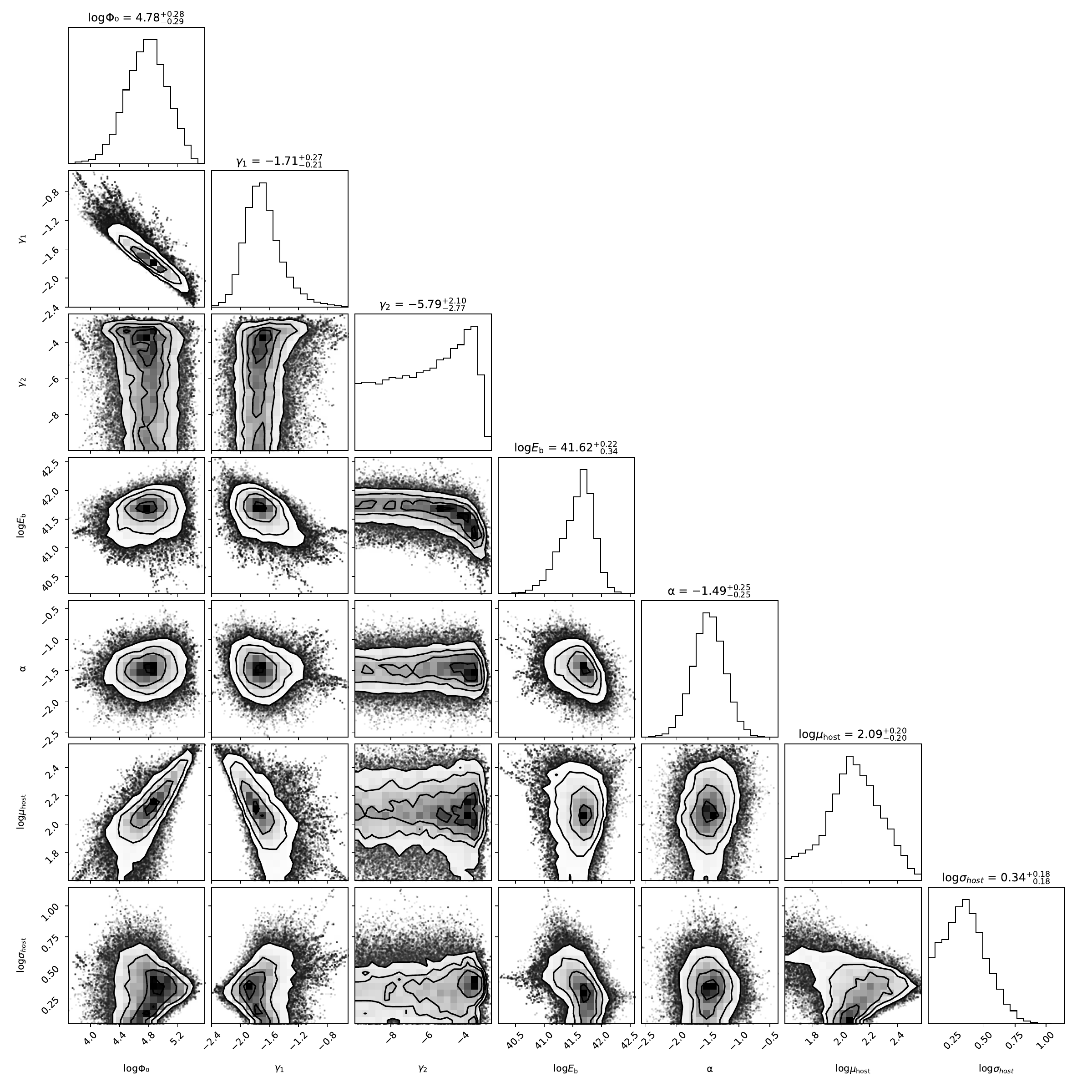}
		\includegraphics[width=0.49\textwidth]{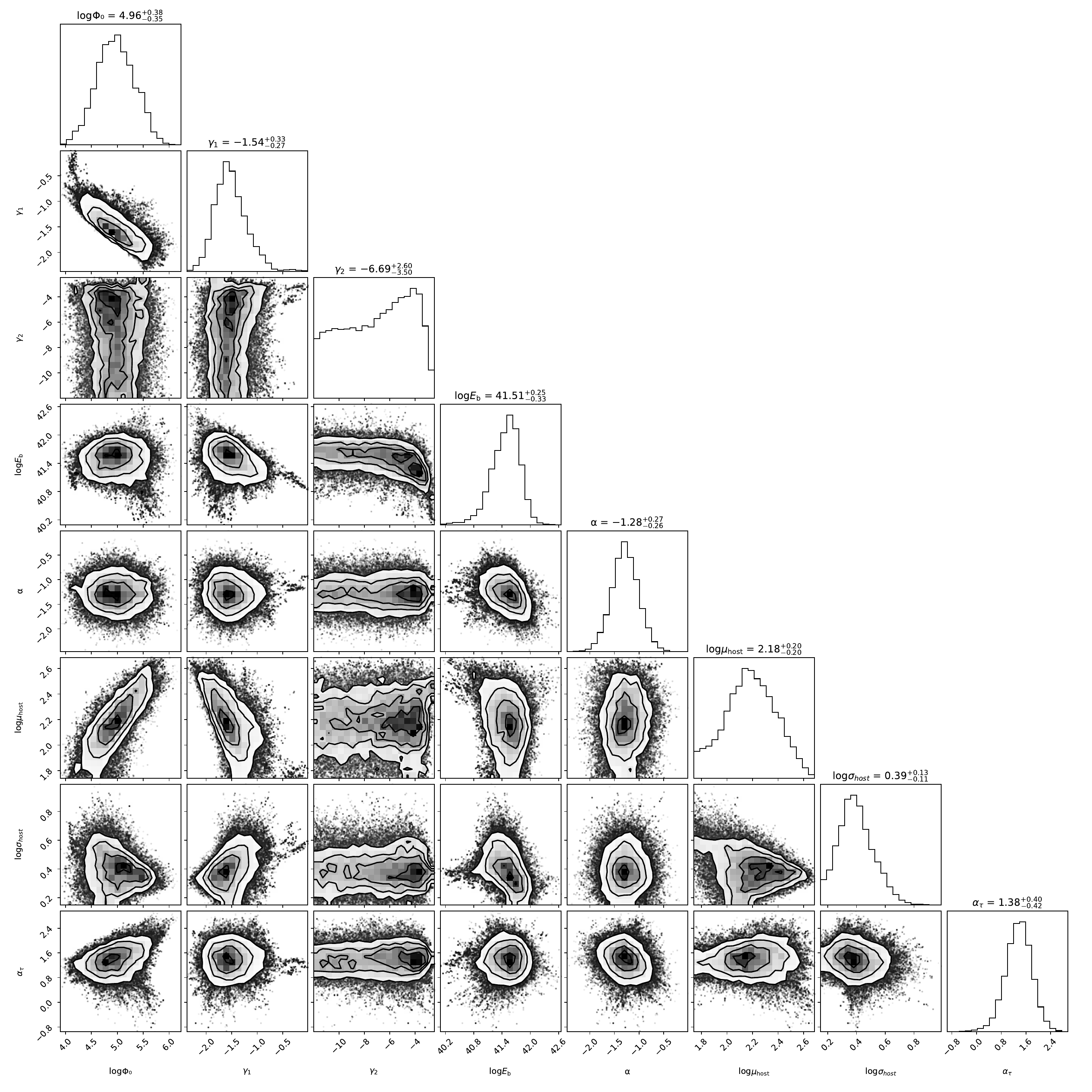}\\
		\caption{Corner plot of the results of the MCMC run for broken power law function. Left panel: redshift model $\Phi(z)\propto \sfr(z)$; right panel: power-law delay model.
			\label{fig:MCMC4}}
	\end{center}
\end{figure*}

\begin{figure*}
	\begin{center}
		\centering 
		\includegraphics[width=0.49\textwidth]{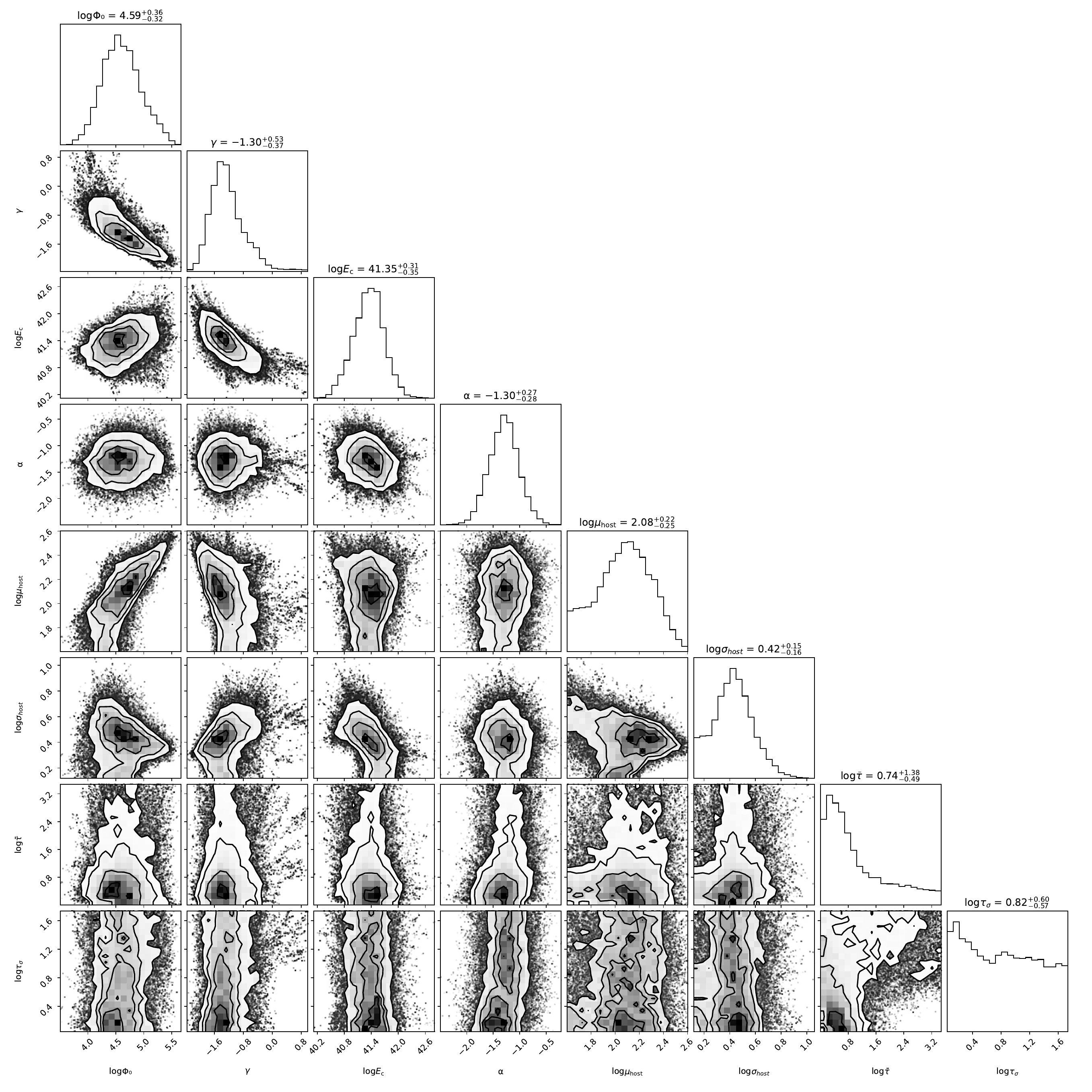}
		\includegraphics[width=0.49\textwidth]{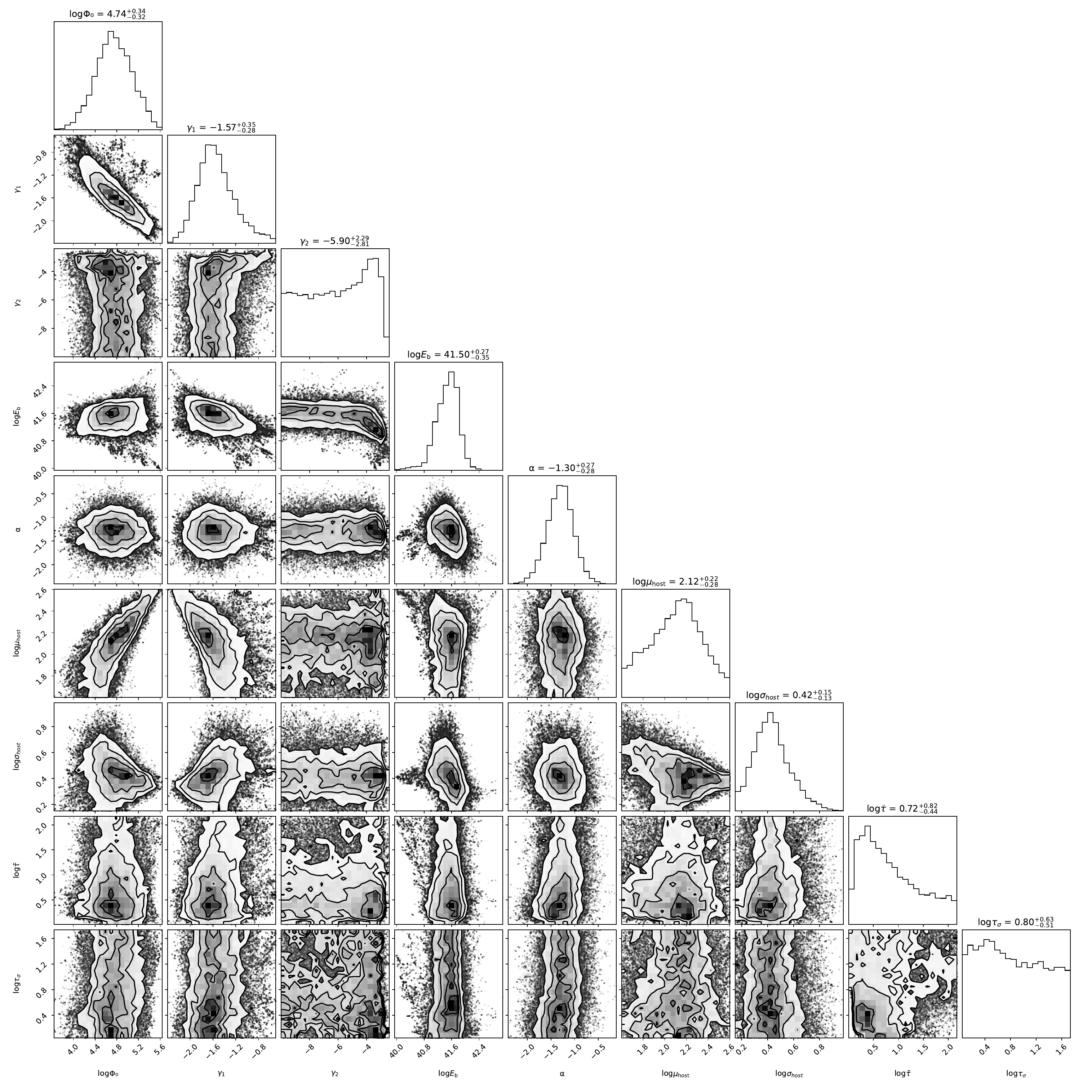}\\
		\caption{Corner plot of the results of the MCMC run. Left panel: log-normal delay model (for Schechter function); right panel: log-normal delay model (for broken power law function).
			\label{fig:MCMC5}}
	\end{center}
\end{figure*}

To ensure robust parameter estimation, we adopt a uniform prior for all parameters except \(\alpha\), for which we use a Gaussian prior of \(\alpha = -1.5 \pm 0.3\) \citep{Macquart2019}. This choice of prior reflects the constraints from previous studies and helps stabilize the fitting process. We employ a Markov Chain Monte Carlo (MCMC) sampler with 30 walkers, 2000 burn-in steps, and 6000 sampling steps. Although convergence is challenging to verify definitively, the similarity between the posterior distributions of the first and second halves of the chains suggests that the chains have converged. The resulting posterior distributions, generated using the \texttt{corner} package \citep{dfm2016_corner}, are shown in Figures~\ref{fig:MCMC1}-\ref{fig:MCMC5}. These figures provide a comprehensive visualization of the parameter space, including correlations between parameters and their uncertainties. The best-fit parameters, represented by the median values of the posterior probability distribution, along with their \(1\sigma\) uncertainties, are summarized in Table~\ref{tab:param_results_2}.

To evaluate the goodness of fit and select the optimal model, we compute the Bayesian Information Criterion (BIC) \citep{Schwarz+1978}, defined as
\begin{eqnarray} 
\label{eqn:BIC} 
	\mathrm{BIC}=-2\mathrm{ln}\mathcal{L}_\mathrm{max} + K\mathrm{ln}(N),
\end{eqnarray}
where \(\mathrm{ln}\mathcal{L}_\mathrm{max}\) is the maximum likelihood estimate, \(K\) is the number of free parameters, and \(N\) is the number of data points. The relative BIC value, \(\Delta \mathrm{BIC}\), is more meaningful than the absolute value and is calculated as
\begin{eqnarray} 
\label{eqn:BIC_1}  
	\Delta \mathrm{BIC} =\mathrm{BIC}_\mathrm{model}-\mathrm{BIC}_\mathrm{reference}.
\end{eqnarray}
Here, the model with the smallest BIC is chosen as the reference model. According to \citet{Jeffreys+1963} and \citet{Liddle2007}, a \(\Delta\mathrm{BIC} \geq 5\) or \(\Delta\mathrm{BIC} \geq 10\) indicates "strong" or "decisive" evidence against a model relative to the reference model. In our analysis, we adopt the SFH model as the reference model and present the \(\Delta \mathrm{BIC}\) values for all models in the last column of Table~\ref{tab:param_results_2}.
The best-fit parameters are summarized in Table~\ref{tab:param_results_2} for the cutoff power law and the broken power law energy distribution models. And figures~\ref{fig:MCMC1}-\ref{fig:MCMC5} specifically present the contour plots of the corresponding fitting results.

Our analysis reveals critical insights into the population properties and cosmological evolution of FRBs through detailed parameter estimation and model comparisons.
For the cutoff power law energy distribution combined with the SFH redshift model—the statistically preferred reference scenario—we derive a local volumetric rate of \(\log(\Phi_0/\mathrm{Gpc}^{-3}\mathrm{yr}^{-1}) = 4.67^{+0.30}_{-0.31}\), a power-law index \(\gamma = -1.49^{+0.37}_{-0.27}\), and a cutoff energy \(\log (E_\mathrm{c}/\mathrm{erg}) = 41.45^{+0.27}_{-0.32}\). The host galaxy DM contributions are tightly constrained to \(\log(\mu_\mathrm{host}/\mathrm{pc \ cm}^{-3}) = 2.07^{+0.18}_{-0.20}\) and \(\log(\sigma_\mathrm{host}/\mathrm{pc \ cm}^{-3}) = 0.37^{+0.17}_{-0.15}\), with spectral index \(\alpha = -1.43^{+0.25}_{-0.27}\). These parameters exhibit relative uncertainties below 15\%, demonstrating robust convergence. 

\begin{figure*}
	\begin{center}
		\centering 
		\includegraphics[width=1\textwidth]{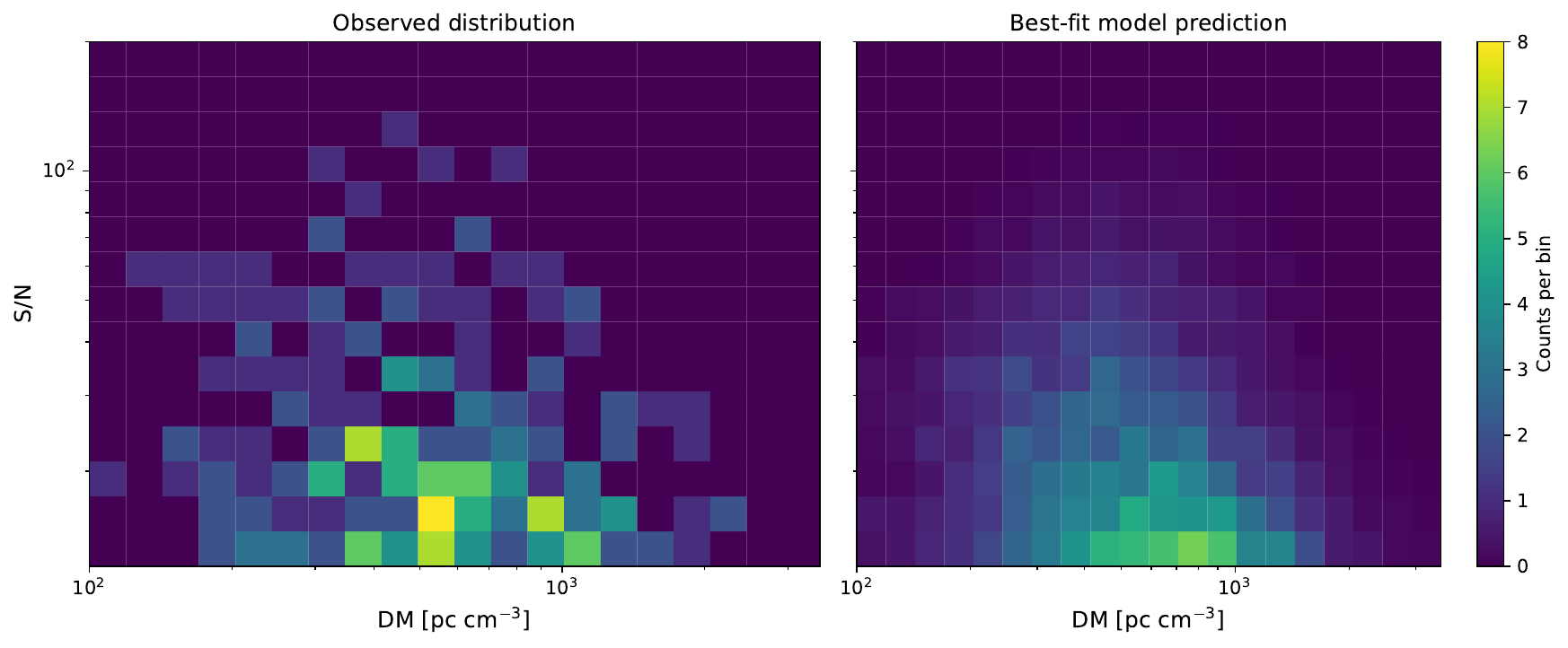}\\
		\caption{ Comparison of the S/N–DM distribution for the observed data (left) and the best-fit model prediction (right). The distributions exhibit qualitative similarity, consistent with expectations derived from Figure~\ref{fig:1}. 
			\label{fig:2}}
	\end{center}
\end{figure*}

To evaluate the goodness-of-fit of our best model—the cutoff power-law energy distribution combined with the SFH redshift distribution—we provide direct visual and statistical comparisons between the model predictions and the  observations. In Figure \ref{fig:1}, we present the one-dimensional marginal distributions of signal-to-noise ratio (S/N) and dispersion measure (DM), comparing the observed data $n_{ij}$ with the corresponding predicted values $\zeta_{ij}$ derived from the posterior distribution of our MCMC analysis. The predicted distributions closely trace the observed data.
To quantify this consistency, we perform Kolmogorov–Smirnov (KS) tests on both the S/N and DM distributions. The resulting p-values are 0.34 and 0.55, respectively, indicating that we cannot reject the null hypothesis that the observed and model-predicted samples are drawn from the same distribution. These values strongly support the conclusion that our best-fit model is a statistically acceptable description of the data.
Furthermore, Figure \ref{fig:2} provides a two-dimensional comparison of the observed and predicted joint distributions in S/N–DM space. The similarity in both the global structure and fine-scale features between the two distributions reinforces the validity of the model. Together, these visual and statistical diagnostics demonstrate that our model not only achieves the lowest BIC value but also provides a good fit to the observed data across both one-dimensional and two-dimensional spaces.

When introducing redshift evolution parameters, all models are statistically disfavored. The density evolution model (\(\beta = 0.11^{+0.91}_{-0.94}\)) shows no significant deviation from zero (\(1\sigma\) range: \(-0.83\)–\(1.02\)), while the energy evolution parameter \(k = 0.08^{+0.32}_{-0.33}\) remains consistent with a non-evolving scenario. The power-law evolution model ($\delta = 2.44^{+0.97}_{-0.89}$) yields results that are consistent with those from the best-fitting SFH model, including a volumetric rate of $\log(\Phi_0/\mathrm{Gpc}^{-3}\mathrm{yr}^{-1}) = 4.69^{+0.36}_{-0.32}$ and an energy spectral index of $\gamma = -1.44^{+0.49}_{-0.35}$. However, it is statistically disfavored relative to the SFH model, with a BIC difference of $\Delta \mathrm{BIC} = 12.29$, indicating that the data prefer the latter. The delay models introducing log-normal and power-law time distributions (\(\mathrm{log}(\bar{\tau}/\mathrm{Gyr}) = 0.74^{+1.38}_{-0.49}\), \(\mathrm{log}(\sigma_\tau/\mathrm{Gyr}) = 0.83^{+0.60}_{-0.57}, \alpha_\tau=1.43^{+0.38}_{-0.37}\) Gyr) fail to improve fits despite added complexity (\(\Delta \mathrm{BIC} = 28.88, \Delta \mathrm{BIC} = 23.59\), respectively), with \(\tau_\sigma\) uncertainties spanning over 200\% of the median value.

The broken power law framework exhibits fundamental instability. For the SFH + broken power law combination, the high-energy slope \(\gamma_2 = -5.79^{+2.10}_{-2.77}\) shows catastrophic uncertainty (68\% CI: \(-8.56\)–\(-3.69\)), while the break energy \(\log (E_\mathrm{b}/\mathrm{erg}) = 41.62^{+0.22}_{-0.34}\)  displays asymmetric errors exceeding 50\%. These issues persist across all redshift models, with \(\gamma_2\) uncertainties expanding further in delay scenarios (Log-normal: \(\gamma_2 = -5.90^{+2.29}_{-2.81}\), Power-law: \(\gamma_2 = -6.69^{+2.60}_{-3.50}\)). Consequently, the broken power law is statistically rejected (\(\Delta \mathrm{BIC} = 13.31\) versus cutoff power law).

The absence of detectable redshift evolution in FRB energetics or densities strongly supports a progenitor population linked to recent star formation. All evolution parameters (\(\beta, k\)) overlap with zero within \(1\sigma\). Combined with the SFH model’s dominance, this suggests FRBs originate from young stellar populations rather than delayed channels like compact binary mergers. The consistency of host galaxy DM parameters (\(\log(\mu_\mathrm{host}/\mathrm{pc \ cm}^{-3}) \approx 2.07\), \(\log(\sigma_\mathrm{host}/\mathrm{pc \ cm}^{-3}) \approx 0.37\)) across models further validates DM-based redshift estimation methods.
 In the next section, we will compare these results with previous studies and discuss the implications of our findings in a broader context.

\section{Discussion \label{sec:Discussion}}
In this study, we analyze the FRB population based on a large sample of bursts from the CHIME/FRB Catalog~1, while carefully considering selection effects,  by using the method developed by \citet{shin+2023}. The redshift, energy, and evolutionary characteristics of the FRB population are constrained. In this section, we discuss the implications of our findings, comparing them with previous results in the field.

\subsection{The Energy Distribution Models \label{subsec:Energy_model}} 
Understanding the energy distribution of FRBs is crucial for shedding light on their origins.
In this study, we investigate two models for the FRB energy distribution: the cutoff power law and the broken power law. As shown in Table~\ref{tab:param_results_2}, the energy distribution parameters are tightly constrained, demonstrating the robustness of our results across different models. However, based on model comparison using the Bayesian Information Criterion, we find that the cutoff power law is statistically favored over the broken power law model. This indicates that the CHIME/FRB data is best described by a simple cutoff power law distribution for FRB energies. In the best-fitting model, which combines the cutoff power law with the SFH redshift model, the characteristic energy cutoff is \(\mathbf{log}(E_\mathrm{c}/\mathrm{erg}) = 41.45^{+0.27}_{-0.32}\) and the power-law index is \(\gamma = -1.49^{+0.37}_{-0.27}\).
It is worth noting that our findings align well with the results of \citet{shin+2023}, who developed the methodology we used and worked with the same FRB sample. 
Additionally, our fitted power-law index is also consistent with the value of \(\gamma = -1.4^{+0.7}_{-0.5}\) reported by \citet{hashimoto+2022}.

However, when comparing our results with other studies using the CHIME/FRB sample, such as those by \citet{2024ApJ...962...73L}, and \citet{2024ApJ...969..123L}, we observe a significant discrepancy. These studies report consistently steeper power-law indices (\(\gamma < -1.8\)), while our results fall within a slightly shallower range. We speculate that this difference primarily to the approach taken to correct for CHIME/FRB’s selection effects. The CHIME FRBs are subject to various biases, such as variations in sensitivity across the sky and frequency, as well as dependence on burst fluence and DM. These selection effects must be thoroughly accounted for to obtain accurate energy distribution estimates. Furthermore, the inability to precisely locate FRBs leads to biases in fluence measurement, which, if not corrected, can distort the inferred energy distribution. We believe that these unaccounted biases in those studies could contribute to the overestimation of the number of weaker FRBs and a misrepresentation of the energy distribution.
Besides, we also note that \cite{james+2022_energy,2025PASA...42....3A} had reported a steeper  power-law index of $\gamma< -1.8$ based on the data from ASKAP. \citet{luo+2020_lumin} investigated the luminosity function based on the Parkes and ASKAP FRBs and found a power-law index of \(\gamma = -1.79^{+0.31}_{-0.35}\), which is also smaller than our result, although it is still marginally consistent within the margin of error.  These deviations might be attributed to the fact that different samples were used.

Furthermore, motivated by the strong redshift evolution in FRB energy reported by \citet{chen+2024_redshift}, who found $E \propto (1+z)^{6.66}$, and similarly noted by \citet{Zhang+2025}, we explore the possibility of redshift evolution in the energy distribution, parameterized as \(E \propto (1+z)^k\). Our results show no significant evidence of redshift evolution in energy,  with \(k = 0.08^{+0.32}_{-0.33}\) for the cutoff power law energy distribution. Our result again stands in contrast to the claims of \citet{chen+2024_redshift,Zhang+2025}. The differences in results may arise from the methodologies employed. In particular, non-parametric techniques like Kendall's \(\tau\) statistics, used in \citet{chen+2024_redshift,Zhang+2025}, are difficult to fully account for the complex selection effects, and even less capable of correcting the biases in FRB fluence measurements. This may very likely lead to an overestimation of the redshift evolution in energy in their results.

\subsection{The Redshift Distribution Models \label{subsec:{Redshift_model}}}

In this study, we systematically explored various redshift distribution models for FRBs, with a particular focus on whether the FRB population follows the cosmic star formation history (SFH). The expectation is that FRBs, potentially originating from young magnetars, might trace the SFH. However, previous studies have largely suggested that the FRB population does not track the SFH directly, instead requiring a time delay of several billion years relative to star formation \citep{hashimoto+2022,qiang+2022,zhang+2022,2024ApJ...962...73L,2024ApJ...969..123L,2025arXiv250115530Z}. In contrast, our results show that the FRB population closely follows the SFH, without the need for a delayed SFH model, which is consistent with the results of \citet{shin+2023, 2025A&A...698A.127M}. 
Again, we believe that the differences between our results and those of previous studies are likely due to the way selection effects are handled. In particular, FRB detection in the low-dispersion DM regime is influenced by radio frequency interference (RFI), which can introduce a significant bias in the observed burst rates. If these biases are not properly accounted for, they may lead to an overestimation of the FRB rate at low redshift \citep{2023PASA...40...57J}. Additionally, assuming a simple flat spectrum for FRBs without considering their spectral index may further exaggerate the burst rate at lower redshifts \citep{james+2022_energy}. These biases could explain why previous studies show a discrepancy, with their redshift distributions deviating from the SFH.

In our study, we also explored alternative redshift models. We repeated the analysis conducted by \citet{shin+2023}, where the redshift distribution follows the form \(\Phi(z) \propto [\sfr(z)]^n\). Our results confirm their conclusion, with \(n = 1.08^{+0.40}_{-0.38}\), which strongly supports the idea that the FRB population tracks the SFH without requiring any significant modification. 
Inspired by studies on gamma-ray bursts, we also investigated whether the FRB redshift distribution might include an additional redshift evolution term, parameterized as \(\Phi(z) \propto \sfr(z) (1+z)^{\beta}\). The GRB population has been shown to exhibit such evolution \citep{2010MNRAS.406..558Q}, and we explored whether a similar form could describe FRBs. Our results indicate \(\beta = 0.11^{+0.91}_{-0.94}\), suggesting that the FRB population does not require such an additional evolution factor. 
{Finally, motivated by the work of \citet{chen+2024_redshift}, who found that the FRB redshift distribution evolves as \(\Phi(z) \propto (1+z)^{ -5}\), we tested this model as well. Surprisingly, we obtained \(\delta = 2.44^{+0.97}_{-0.89}\), which is in contrast to their findings, as our result points to a positive evolution rather than a negative one. 
However, this model exhibited a significantly inferior goodness-of-fit to the data relative to the SFH model, as evidenced by a Bayesian information criterion difference of 12.29. Therefore, the model was conclusively rejected.}
As discussed previously, we suspect that the differences between our results and those of \citet{chen+2024_redshift} and \citet{Zhang+2025} may stem from their failure to fully account for selection effects and the biases in fluence measurements. These uncorrected biases likely lead to exaggerated redshift evolution in their analyses.

To sum up, our results suggest that the FRB population closely tracks the SFH, with no significant redshift evolution required. We argue that previous studies showing a significant delay or strong redshift evolution may have been affected by observational biases. Further refinement in the measurement of FRB redshifts, especially through the accurate localization of host galaxies, will likely improve our understanding of the redshift evolution of the FRB population.

\section{Conclusions \label{sec:Conclusions}}

In this study, we have presented a comprehensive analysis of FRBs using the CHIME/FRB Catalog~1, focusing on their energy and redshift distributions. We systematically explored various models to better understand the underlying physical processes driving FRB phenomena, with a particular emphasis on their population characteristics and cosmological evolution.

We started by considering two energy distribution models: the cutoff power-law and the broken power-law. Our results show that the cutoff power law model best describes the FRB population, with a characteristic energy cutoff at $E_\mathrm{c} = 2.82^{+2.43}_{-1.47} \times 10^{41}$ erg and a power-law index $\gamma = -1.49^{+0.37}_{-0.27}$.  This is consistent with previous studies that fully account for observational biases, particularly the selection effects inherent in the CHIME survey.  Importantly, our analysis reveals no significant evidence for redshift evolution in the FRB energy function, which is in contrast to some previous studies that suggested stronger redshift evolution.

In terms of redshift distributions, we found that the FRB population follows the SFH model, with no compelling evidence for the time delays proposed in some previous studies. This finding underscores the possibility that FRBs are primarily generated by young magnetars, consistent with the SFH of the universe, and suggests that there is no need to invoke significant delays between the FRB population and SFH.  We further demonstrated that the SFH model, in its simplest form, adequately describes the redshift evolution of FRBs without additional evolutionary factors.


In conclusion, our findings underscore the importance of correcting for selection effects in FRB population studies. While the SFH-tracking model remains the most plausible explanation for the redshift evolution of FRBs, further progress will depend on future advancements in observational data. Larger samples with precise redshifts of host galaxies and accurate localizations will be crucial for refining our understanding of FRB evolution and energy scales.
Future catalogs from CHIME/FRB and next-generation telescopes will enable more robust tests of FRB origins and their potential as cosmological probes.

\section*{acknowledgments}
We thank Kaitlyn Shin for useful discussions. This work is supported by the National Natural Science Foundation of China (grant No.12203013) and the GuangXi Science Foundation (grant No 2023GXNSFBA026030).  C.M.D. is partially supported by the National Natural Science Foundation of China (grant No. 12494575).

\bibliography{references}{}
\bibliographystyle{aasjournal}

\begin{rotatetable}
	\begin{flushleft} 
		\small  
		\begin{longrotatetable}
			\begin{deluxetable*}{lccccccccccccccc}
				\centering
				\tablecaption{Best-fit parameters for different models.}
				\label{tab:param_results_2}
				\tablehead{\colhead{Redshift} & \colhead{evolution} &\colhead{log$\Phi_0$}& \colhead{$\gamma^{a}$}&\colhead{$\gamma_2$} &\colhead{log$E^{b}_\mathrm{c}$} &\colhead{$\alpha$} &\colhead{log$\mu_\mathrm{host}$ }& \colhead{log$\sigma_\mathrm{host}$} & \colhead{log$\bar{\tau}$}  & \colhead{log$\tau_\sigma$}&\colhead{$\alpha_{\tau}$}& \colhead{$\Delta \mathrm{BIC}$}\\ \colhead{Model$(\Phi(z)\propto)$}&\colhead{parameter}}
				\startdata
				\multicolumn{13}{c}{\textbf{Schechter function}} \\[0.6em]
				$\sfr(z)$ &{-}& $4.67^{+0.30}_{-0.31}$ &$-1.49^{+0.37}_{-0.27}$ &{-}& $41.45^{+0.27}_{-0.32}$  & $-1.43^{+0.25}_{-0.27}$ & ${2.07^{+0.18}_{-0.20}}$ &${0.37^{+0.17}_{-0.15}}$ &{-}&{-} &{-}&$0$ (reference) \\[0.5em]
				$\sfr_\mathrm{d_{L}}(z)^d$ & {-} & 4.59$^{+0.36}_{-0.32}$ & $-1.30^{+0.53}_{-0.37}$ &{-}&$41.35^{+0.31}_{-0.35}$ & $-1.30^{+0.27}_{-0.28}$  &  ${2.08^{+0.22}_{-0.25}}$&${0.42^{+0.15}_{-0.16}}$  &${0.74^{+1.38}_{-0.49}}^{c}$&${0.82^{+0.60}_{-0.57}}^{c}$&{-}&$28.88$\\[0.5em]
				$\sfr_\mathrm{d_{P}}(z)^e$ & {-} & 4.75$^{+0.42}_{-0.37}$ & $-1.11^{+0.66}_{-0.45}$ &{-}&$41.24^{+0.32}_{-0.35}$ & $-1.30^{+0.28}_{-0.27}$  &  ${2.14^{+0.20}_{-0.22}}$&${0.41^{+0.14}_{-0.12}}$  &{-}&{-}&${1.43^{+0.38}_{-0.37}}$&$23.59$\\[0.5em]
				$[\sfr(z)]^n$ & $n=1.08^{+0.40}_{-0.38}$ & $4.71^{+0.32}_{-0.31}$ &$-1.54^{+0.39}_{-0.32}$ &{-}& $41.47^{+0.28}_{-0.31}$  & $-1.48^{+0.29}_{-1.29}$ & ${2.08^{+0.21}_{-0.24}}$&${0.38^{+0.16}_{-0.14}}$  &{-}&{-} &{-}&$12.10$\\[0.5em]
				$\sfr(z)(1+z)^{\beta}$ & ${\beta}=0.11^{+0.91}_{-0.94}$ & $4.68^{+0.32}_{-0.31}$ &$-1.51^{+0.44}_{-0.32}$&{-} & $41.45^{+0.29}_{-0.33}$  & $-1.48^{+0.30}_{-0.29}$ & ${2.08^{+0.20}_{-0.20}}$&${0.37^{+0.17}_{-0.16}}$  &{-} &{-}&{-}&$12.12$ \\[0.5em]
				$(1+z)^\delta$& $\delta$=$2.44^{+0.97}_{-0.89}$ & $4.69^{+0.36}_{-0.32}$ &$-1.44^{+0.49}_{-0.35}$ &{-}& $41.39^{+0.28}_{-0.35}$  & $-1.51^{+0.29}_{-0.29}$ & ${2.11^{+0.20}_{-0.20}}$&${0.39^{+0.15}_{-0.15}}$  &{-} &{-}&{-}& $12.29$ \\[0.6em]\hline
				\multicolumn{13}{c}{\text{ FRB energy evolve with redshift ($E\propto(1+z)^k$ ) }} \\[0.6em]
				$\sfr(z)$ & $k=0.08^{+0.32}_{-0.33}$ & $5.40^{+0.62}_{-0.51}$ &$-1.48^{+0.34}_{-0.28}$ &{-}& $41.43^{+0.28}_{-0.30}$  & $-1.51^{+0.30}_{-0.29}$ &${2.06^{+0.21}_{-0.22}}$&${0.38^{+0.17}_{-0.18}}$  &{-} &{-}&{-}&$12.09$\\[0.5em]\hline
				\multicolumn{13}{c}{\textbf{Broken power law function }} \\[0.6em]
				$\sfr(z)$&{-}& $4.78^{+0.28}_{-0.29}$ &${-1.71^{+0.27}_{-0.21}}$& ${-5.79^{+2.10}_{-2.77}}^{c}$& $41.62^{+0.22}_{-0.34}$  & $-1.49^{+0.25}_{-0.25}$ & ${2.09^{+0.20}_{-0.20}}$&${0.34^{+0.18}_{-0.18}}$  &{-}&{-}&{-}&$13.31$ \\[0.5em]
				$\sfr_\mathrm{d_{L}}(z)^d$& {-} & $4.74^{+0.34}_{-0.32}$ & ${-1.57^{+0.35}_{-0.28}}$&	${-5.90^{+2.29}_{-2.81}}^{c}$ &$41.50^{+0.27}_{-0.35}$ & $-1.30^{+0.27}_{-0.28}$  &  ${2.12^{+0.22}_{-0.28}}$ &${0.42^{+0.15}_{-0.13}}$ &${0.72^{+0. 82}_{-0.44}}^{c}$ &${0.80^{+0.63}_{-0.51}}^{c}$ &{-} &$41.43$\\[0.5em]
				$\sfr_\mathrm{d_{P}}(z)^e$& {-} & $4.96^{+0.38}_{-0.35}$ & ${-1.54^{+0.33}_{-0.27}}$&	${-6.69^{+2.60}_{-3.50}}^{c}$ &$41.51^{+0.25}_{-0.33}$ & $-1.28^{+0.27}_{-0.26}$  &  ${2.18^{+0.20}_{-0.20}}$ &${0.39^{+0.13}_{-0.11}}$&{-}&{-} &${1.38^{+0.40}_{-0.42}}$ &$33.56$\\[0.5em]
				\enddata
				\tablenotetext{} {Notes. The best-fit results are represented by the medians of the posterior distributions, with the errors reflecting the central 68$\%$ of the samples. The parameter $\Phi_0$ is in units of Gpc$^{-3}$yr$^{-1}$, the parameter $E_\mathrm{c} (E_\mathrm{b})$ is in units of erg, the parameters $\mu_\mathrm{host}$ and $\sigma_\mathrm{host}$ are in units of pc cm$^{-3}$, the parameters $\overline\tau$, $\sigma_\tau$ and $\alpha_{\tau}$ are in units of billion years (Gyr). }
				\tablenotetext{a} {For the broken power law function, $\gamma$ here corresponds to $\gamma_1$ in Eq.(\ref{eqn:broken_power_law_function}).} 
				\tablenotetext{b} {For the broken power law function, $E_\mathrm{c}$ here corresponds to $E_\mathrm{b}$ in Eq.(\ref{eqn:broken_power_law_function}).} 
				\tablenotetext{c}{Note that these parameters fail to converge in the current model.}
				\tablenotetext{d}{Note that $\sfr_\mathrm{d_{L}}(z)$ refers to the log-normal delay model.}
				\tablenotetext{e}{Note that $\sfr_\mathrm{d_{P}}(z)$ refers to the power-law delay model.}
			\end{deluxetable*}
		\end{longrotatetable}
	\end{flushleft}
\end{rotatetable}


\end{document}